\documentclass[useAMS,usenatbib]{mn2e}

\usepackage{pdflscape}
\usepackage{epstopdf}
\usepackage{graphicx}
\usepackage{textcomp}
\usepackage{courier}
\usepackage{caption}
\usepackage{amssymb,amsmath}
\usepackage{url}
\usepackage{deluxetable}
\usepackage{gensymb}
\usepackage{tabularx}

\newcommand{\aj}{AJ}
\newcommand{\apj}{ApJ}
\newcommand{\apjl}{ApL}
\newcommand{\apjs}{ApJS}

\newcommand{\mnras}{MNRAS}
\newcommand{\araa}{ARA\&A}
\newcommand{\nat}{Nature}
\newcommand{\pasp}{PASP}
\newcommand{\aap}{A\&A}
\newcommand{\procspie}{SPIE}
\newcommand{\pasj}{PASJ}

\title[SN2012ab: A Peculiar Type IIn Supernova]{SN2012ab: A Peculiar Type IIn Supernova with Aspherical Circumstellar Material}
\author[Bilinski et. al.]{Christopher Bilinski$^{1}$\thanks{E-mail:
cgbilinsk@gmail.com}, Nathan Smith$^{1}$, G. Grant Williams$^{2}$, Paul Smith$^{1}$, 
\newauthor WeiKang Zheng$^{3}$, Melissa L. Graham$^{3,4}$, Jon C. Mauerhan$^{3}$, Jennifer E. Andrews$^{1}$ 
\newauthor Alexei V. Filippenko$^{3,5}$, Carl Akerlof$^{6}$, E. Chatzopoulos$^{7}$, Jennifer L. Hoffman$^{8}$
\newauthor Leah Huk$^{8}$, Douglas C. Leonard$^{9}$, G. H. Marion$^{10}$, Peter Milne$^{1}$
\newauthor Robert M. Quimby$^{11}$, Jeffrey M. Silverman$^{10}$, Jozsef Vink\'{o}$^{12,10,13}$, J. Craig Wheeler$^{10}$,
\newauthor Fang Yuan$^{6,14,15}$
\\
$^{1}$Steward Observatory, University of Arizona, 933 N. Cherry Avenue, Tucson, AZ 85721, USA\\
$^{2}$MMT Observatory, Tucson, AZ 85721-0065, USA\\
$^{3}$Department of Astronomy, University of California, Berkeley, CA 94720-3411, USA\\
$^{4}$Department of Astronomy, Box 351580, U.W., Seattle, WA 98195-1580, USA\\
$^{5}$Miller Senior Fellow, Miller Institute for Basic Research in Science, University of California, Berkeley, CA  94720, USA\\
$^{6}$Department of Physics, University of Michigan, 450 Church Street, Ann Arbor, MI 48109-1040, USA\\
$^{7}$Department of Physics and Astronomy, Louisiana State University, Baton Rouge, LA 70803, USA \\
$^{8}$Department of Physics \& Astronomy, University of Denver, 2112 East Wesley Avenue, Denver, CO 80208, USA \\
$^{9}$Department of Astronomy, San Diego State University, San Diego, CA 92812, USA\\
$^{10}$University of Texas, 1 University Station C1400, Austin, TX 78712-0259, USA\\
$^{11}$Kavli IPMU, The University of Tokyo, 5-1-5 Kahiwanoha, Kashiwa-shi, Chiba, 277-8583, Japan\\
$^{12}$Konkoly Observatory, Research Centre for Astronomy and Earth Sciences, Hungarian Academy of Sciences, Konkoly Thege M. ut 15-17, Budapest, 1121 Hungary\\
$^{13}$Department of Optics and Quantum Electronics, University of Szeged, D\'{o}m t\'{e}r 9, Szeged, 6720 Hungary\\
$^{14}$Research School of Astronomy and Astrophysics, Australian National University, Weston Creek ACT 2611, Australia\\
$^{15}$ARC Centre of Excellence for All-sky Astrophysics}
\begin{document}

\date{Accepted 0000. Received 0000; in original form 0000}

\pagerange{\pageref{firstpage}--\pageref{lastpage}} \pubyear{2017}

\maketitle

\label{firstpage}

\begin{abstract}
We present photometry, spectra, and spectropolarimetry of supernova (SN) 2012ab, mostly obtained over the course of $\sim 300$ days after discovery.  SN~2012ab was a Type IIn (SN~IIn) event discovered near the nucleus of spiral galaxy 2MASXJ12224762+0536247.  While its light curve resembles that of SN~1998S, its spectral evolution does not.  We see indications of CSM interaction in the strong intermediate-width emission features, the high luminosity (peak at absolute magnitude $M=-19.5$), and the lack of broad absorption features in the spectrum.  The H$\alpha$ emission undergoes a peculiar transition.  At early times it shows a broad blue emission wing out to $-14{,}000$ km $\mathrm{s^{-1}}$ and a truncated red wing.  Then at late times ($>$ 100\,days) it shows a truncated blue wing and a very broad red emission wing out to roughly $+20{,}000$ km $\mathrm{s^{-1}}$.  This late-time broad red wing probably arises in the reverse shock.  Spectra also show an asymmetric intermediate-width H$\alpha$ component with stronger emission on the red side at late times.  The evolution of the asymmetric profiles requires a density structure in the distant CSM that is highly aspherical.  Our spectropolarimetric data also suggest asphericity with a strong continuum polarization of $\sim 1$--3\% and depolarization in the H$\alpha$ line, indicating asphericity in the CSM at a level comparable to that in other SNe~IIn.  We estimate a mass-loss rate of $\dot{M} = 0.050\, {\rm M}_{\odot}\,\mathrm{yr^{-1}}$ for $v_{\rm pre} = 100$\,km\,$\mathrm{s^{-1}}$ extending back at least 75\,yr prior to the SN.  The strong departure from axisymmetry in the CSM of SN~2012ab may suggest that the progenitor was an eccentric binary system undergoing eruptive mass loss.
\end{abstract}


\section{Introduction}
\label{sec:Int}
Type IIn supernovae (SNe~IIn) are known for their strong narrow or intermediate-width hydrogen emission lines superimposed on an otherwise smooth blue continuum \citep{1990MNRAS.244..269S,1997ARA&A}.  These prominent lines likely originate in dense circumstellar material (CSM) that was ejected shortly before the SN explosion itself (see \citealp[]{2014ARA&A..52..487S} for a review of pre-SN mass loss).  Spectra of SNe~IIn often show signs of asphericity in the CSM and SN ejecta (SN 1988Z: \citealt{1994MNRAS.268..173C}; SN 1995N: \citealt{2002ApJ...572..350F}; SN 1997eg: \citealt{2008ApJ...688.1186H}; SN 1998S: \citealt{2000ApJ...536..239L,2001ApJ...550.1030W,2005ApJ...622..991F}; SN 2005ip: \citealt{2009ApJ...695.1334S,2014ApJ...780..184K}; SN 2006jd: \citealt{2012ApJ...756..173S}; SN 2006tf: \citealt{2008ApJ...686..467S};  SN 2009ip: \citealt{2014MNRAS.442.1166M,2017arXiv170108885R}; SN 2010jl: \citealt{2012AJ....143...17S,2014ApJ...797..118F}; PTF11iqb: \citealt{2015MNRAS.449.1876S}).  While blueshifted line profiles can be caused by dust formation or occultation by the SN photosphere, other types of line-profile asymmetries and polarization favour real geometrical asphericities.  When indicative of real asphericity, line-profile shapes may be caused by an aspherical explosion (which may tell us about the explosion mechanism), asphericity in the CSM (providing clues to the nature of the progenitor system and its unstable mass loss), or both.  Since SNe~IIn sweep through CSM with time, and because the optical depth changes with time, monitoring the time-evolution of SNe~IIn at many different epochs helps disentangle the various potential sources of asymmetric line emission.

Observational evidence suggests that SNe~IIn are aspherical and may have high polarization signals.  A $\sim 20\%$ level of SN asphericity may result in a detectable 1\% linear polarization signal \citep{1991A&A...246..481H,2005ASPC..342..330L}.  While a number of efforts have been made to explain core-collapse SNe in terms of axisymmetric jets \citep{1999ApJ...524L.107K,2002ApJ...568..807W,2002ApJ...579..671W}, observational evidence in the form of loops in the Q/U plane suggest that even these axisymmetric models may not be sufficient for all types of core-collapse SNe \citep{2008ApJ...688.1186H,2008ARA&A..46..433W,2009ApJ...705.1139M}.  In contrast to SNe~IIn, SNe~II-P generally have shown very low levels of polarization at early times (\citealp{2001PASP..113..920L,2002AJ....124.2490L}; however, see \citealp{2016IAUFM..29B.458L,2017ApJ...834..118M}).  The initially low polarization levels often rise during the plateau phase (e.g., \citealp{2016IAUFM..29B.458L}), with a polarization angle that typically remains nearly fixed throughout (e.g., \citealp{2001ApJ...553..861L,2017ApJ...834..118M}).  Occasionally, a sharp rise in the polarization signal is seen during the transition to the nebular phase \citep{2006Natur.440..505L,2010ApJ...713.1363C}, perhaps suggesting that the core of the SN is more aspherical than the early-time photosphere \citep{2005AstL...31..792C,2006AstL...32..739C}.  However, as demonstrated by the modeling of \citet{2011MNRAS.410.1739D}, it is also possible that even large asymmetries during the plateau phase will produce very little polarization, owing to the high optical depth to electron scattering and the fact that geometric information is lost due to multiple scatters.  The ``spike'' that is sometimes seen during the drop off of the plateau may, therefore, be more of an optical-depth effect (i.e., the ``spike'' occurs when $\tau_{e^-}$ has decreased to unity) than a demonstration of increasing asphericity with depth in the atmosphere \citep{2012AIPC.1429..204L}.  The picture for SNe~IIn, on the other hand, is not as well understood.  The primary reason for this is that an effective model for a SN~IIn must not only account for the geometry of the SN ejecta, but also the geometry of the CSM interaction region \citep{2001MNRAS.326.1448C}.  In such cases, the temporal evolution that multi-epoch spectropolarimetry provides becomes particularly important in establishing a physical model.

Spectropolarimetry provides unique insight into these explosions in that it allows us to view the asphericities in both the explosion itself and the recently ejected CSM through a perspective not provided by total-flux spectra alone.  Having multiple epochs of spectropolarimetry allows us to further track these asphericities as the SN evolves.  Shifts in the magnitude and angle of the polarization signal over time reveal different geometries for the CSM shell and the SN ejecta in some SN~IIn spectropolarimetry (e.g., SN 2009ip: \citealt{2014MNRAS.442.1166M,2017arXiv170108885R}; SN 1997eg: \citealt{2008ApJ...688.1186H}).  Shifts in the polarization signal across particular emission or absorption lines tell us about the geometry of the regions from which they arise (Maund et al. 2007).  

SN~2012ab was discovered coincident with the nucleus of the spiral galaxy 2MASXJ12224762+0536247 (redshift $z=0.018$) as a part of the Robotic Optical Transient Search Experiment (ROTSE) at an unfiltered apparent magnitude of $15.8$ ($M = -19.0$\,mag) on 2012 January 31.35 (UT dates are used throughout this paper) with the 0.45-m ROTSE-IIIb telescope at the McDonald Observatory \citep{2012CBET.3022....1V}.  A finder chart SN~2012ab is shown in Figure \ref{fig:SN2012abfinder}.  The object was not detected in an image that was taken two nights earlier on Jan. 29.17 with a limit of $m \approx 18.7$\,mag (absolute $M \approx -16.1$\,mag), indicating that SN~2012ab either exploded earlier and brightened suddenly owing to CSM interaction or was found within $\sim 2$ days of explosion.  It peaked at an unfiltered apparent magnitude of $15.3$ (absolute $M=-19.5$) on February 27.28.  It was reportedly located at $\alpha\mathrm{(J2000)} = 12^\mathrm{h}22^\mathrm{m}47^\mathrm{s}.6$, $\delta\mathrm{(J2000)} = +05\degree36^{\prime}25^{\prime\prime}.0$ \citep{2012CBET.3022....1V}.  We obtain the Milky Way extinction along the line of sight of $A_R=0.045\,\mathrm{mag}$, $A_V=0.057\,\mathrm{mag}$ ($E_{B-V} = 0.018$\,mag; \citealp{2011ApJ...737..103S}), and a redshift-based distance [which assumes $H_0=73\,\mathrm{km\,s^{-1}\,Mpc^{-1}}$ \citep{2005ApJ...627..579R} and takes into account influences from the Virgo cluster, the Great Attractor, and the Shapley supercluster] of $82.3\pm 5.8\,\mathrm{Mpc}$ from the NASA/IPAC Extragalactic Database\footnote{The NASA/IPAC Extragalactic Database (NED) is operated by the Jet Propulsion Laboratory, California Institute of Technology, under contract with the National Aeronautics and Space Administration (NASA; \url{http://ned.ipac.caltech.edu}).}. 

A spectrum of SN~2012ab acquired with the 9.2-m Hobberly Eberly Telescope/Marcario Low-Resolution Spectrograph on February 7.34 revealed a hot, blue continuum with narrow H$\alpha$ and [O \textsc{iii}] emission lines from the host galaxy \citep{2012CBET.3022....1V}.  The redshift calculated from the narrow features in the SN agrees with the Sloan Digital Sky Survey (SDSS) galaxy redshift of 0.018.  A later spectrum acquired on February 16.31 shows broader Balmer emission features superimposed on the blue continuum \citep{2012CBET.3022....1V}.  Because of its narrow H$\alpha$ emission lines, SN~2012ab was classified as a normal SN~IIn.  Here, we present results for SN~2012ab based on two epochs of spectropolarimetry, twelve other epochs of spectroscopy, and early-time photometry.

\section{Observations}
\label{sec:Obs}

\subsection{Photometry}
\begin{figure*}
\centering
\includegraphics[width=1\textwidth,clip=true,trim=0cm 0cm 0cm 0cm]{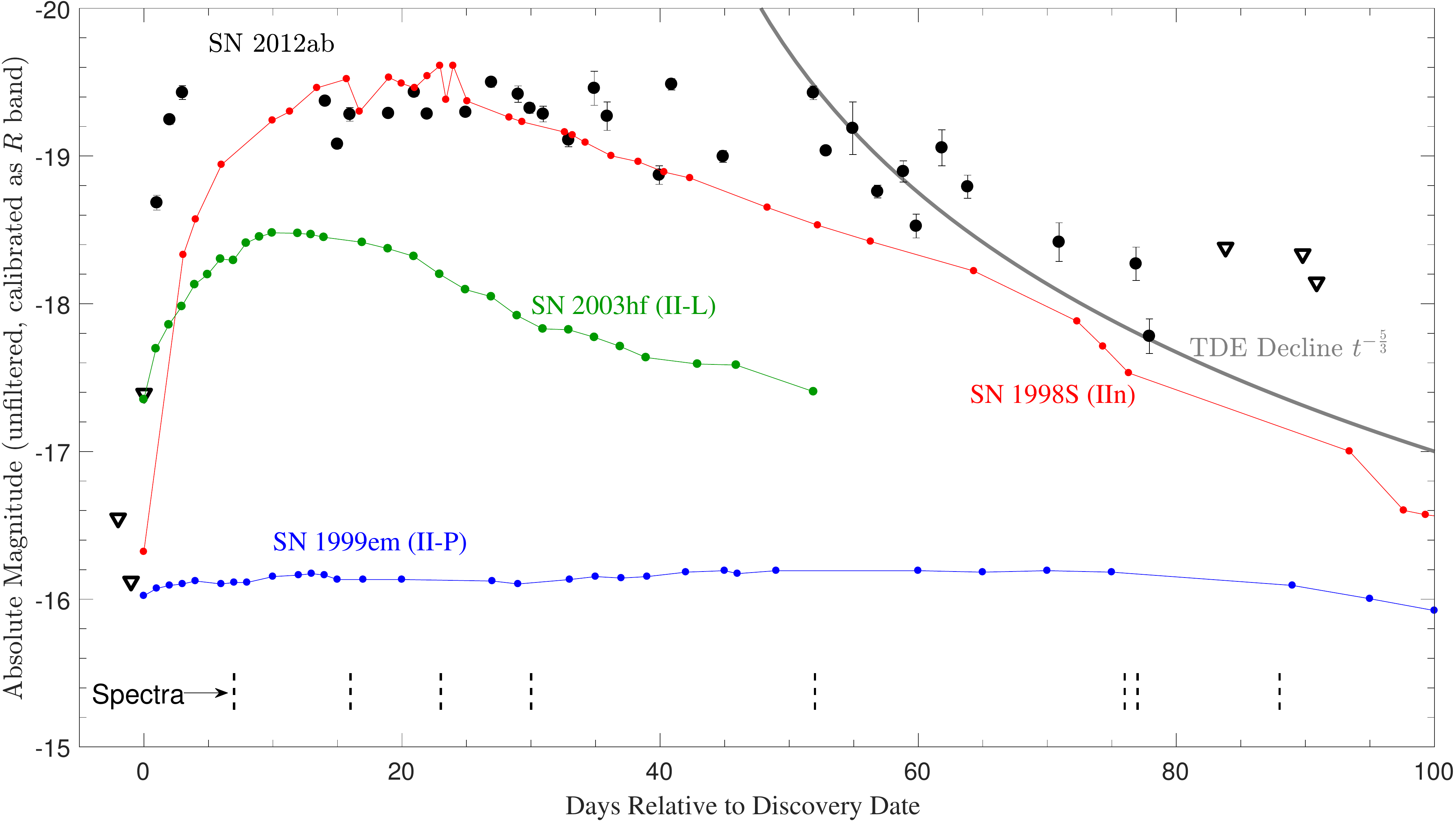}
\caption{ROTSE-IIIb unfiltered absolute magnitudes (calibrated as $\sim R$ band) of SN~2012ab are shown in black circles, corrected for $A_R = 0.24$ mag (see \S 3.1) and $m-M = 34.57$\,mag.  Unfilled black triangles indicate limiting magnitudes.  Dotted vertical black lines at the bottom indicate dates on which we obtained spectra.  For comparison, we have included the light curves of the Type II-P SN~1999em (blue; \citealp{2002PASP..114...35L}), the Type II-L SN~2003hf (green; \citealp{2014MNRAS.445..554F}), the Type IIn SN~1998S (red; \citealp{2000MNRAS.318.1093F,2011arXiv1109.0899P}), and the expected decline rate for a TDE (grey).}
\label{fig:ROTSELightCurve}
\end{figure*}

After the discovery of SN~2012ab, ROTSE-IIIb continued to gather unfiltered data for $\sim100$ days on an almost daily basis.  The ROTSE-IIIb images were bias-subtracted and flat-fielded by an automated
pipeline.  In order to remove contamination from the underlying host galaxy, we constructed a pre-SN galaxy template from images taken in March 2008.  We performed image subtraction \citep{2008ApJ...677..808Y} before running our custom RPHOT photometry program \citep{2006PhDT........13Q}, which was based on the DAOPHOT \citep{1987PASP...99..191S} point-spread-function (PSF) fitting photometry package. All ROTSE-IIIb unfiltered magnitudes have been converted to $R$-band magnitudes via USNO-B1.0 photometric calibrations and included in Table \ref{tab:ROTSELightCurve}.  We also display this ROTSE-IIIb light curve in Figure \ref{fig:ROTSELightCurve} alongside the light curves of another SN~IIn (SN~1998S), a SN~II-P (SN~1999em), a SN~II-L (SN~2003hf), and a theoretical tidal disruption event (TDE).  Image subtraction was complicated by the large pixel scale and the SN being coincident with its host-galaxy nucleus.  The statistical uncertainties (derived from background noise) in Figure \ref{fig:ROTSELightCurve} likely underestimate the actual errors in the photometry.  The photometric scatter in the data is also higher than normal for ROTSE-IIIb data, probably because SN~2012ab is located in the nucleus of its host galaxy.

\begin{figure*}
\centering
\includegraphics[width=0.4\textwidth,clip=true,trim=0cm 0cm 0cm 0cm]{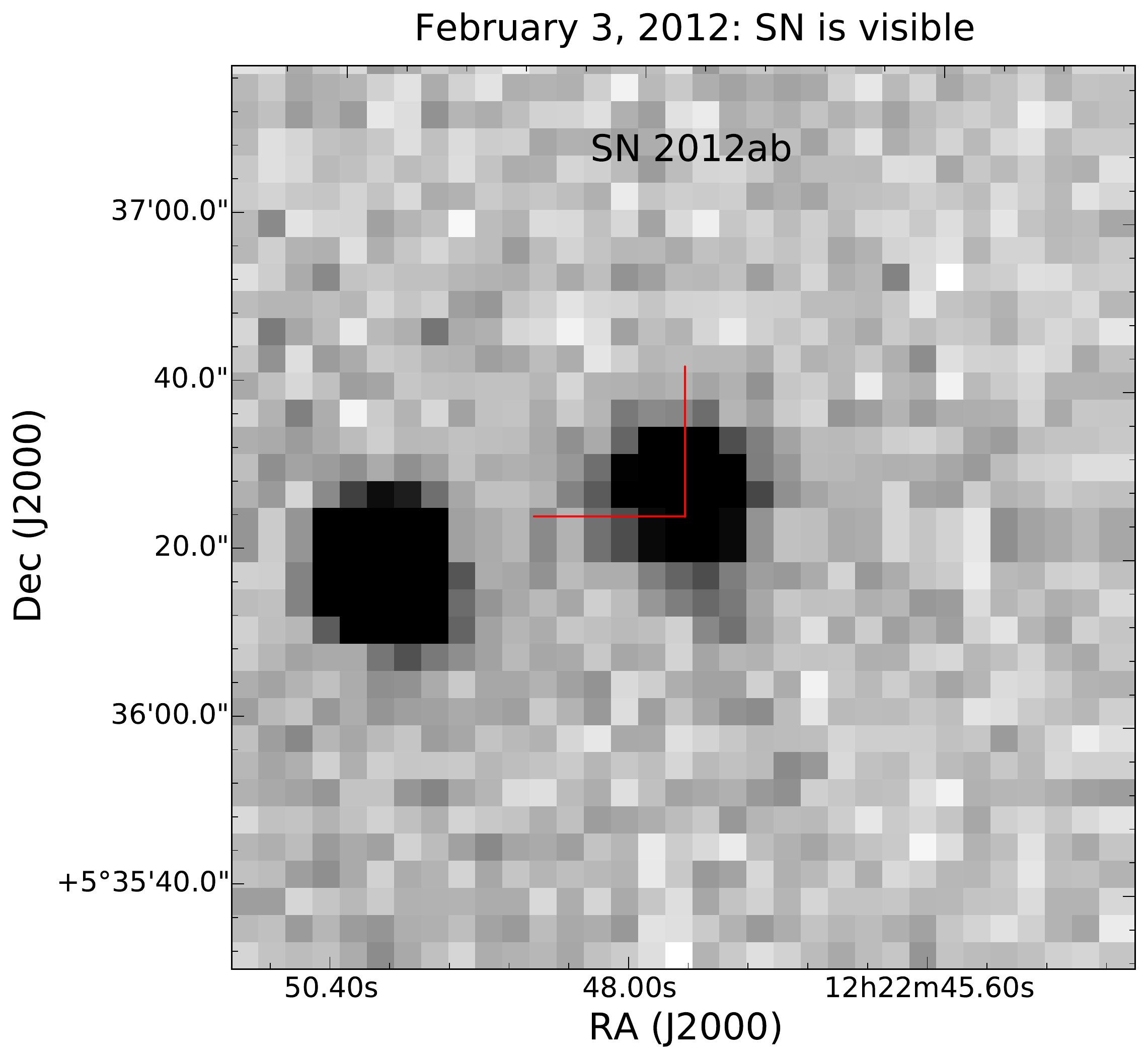}
\includegraphics[width=0.4\textwidth,clip=true,trim=0cm 0cm 0cm 0cm]{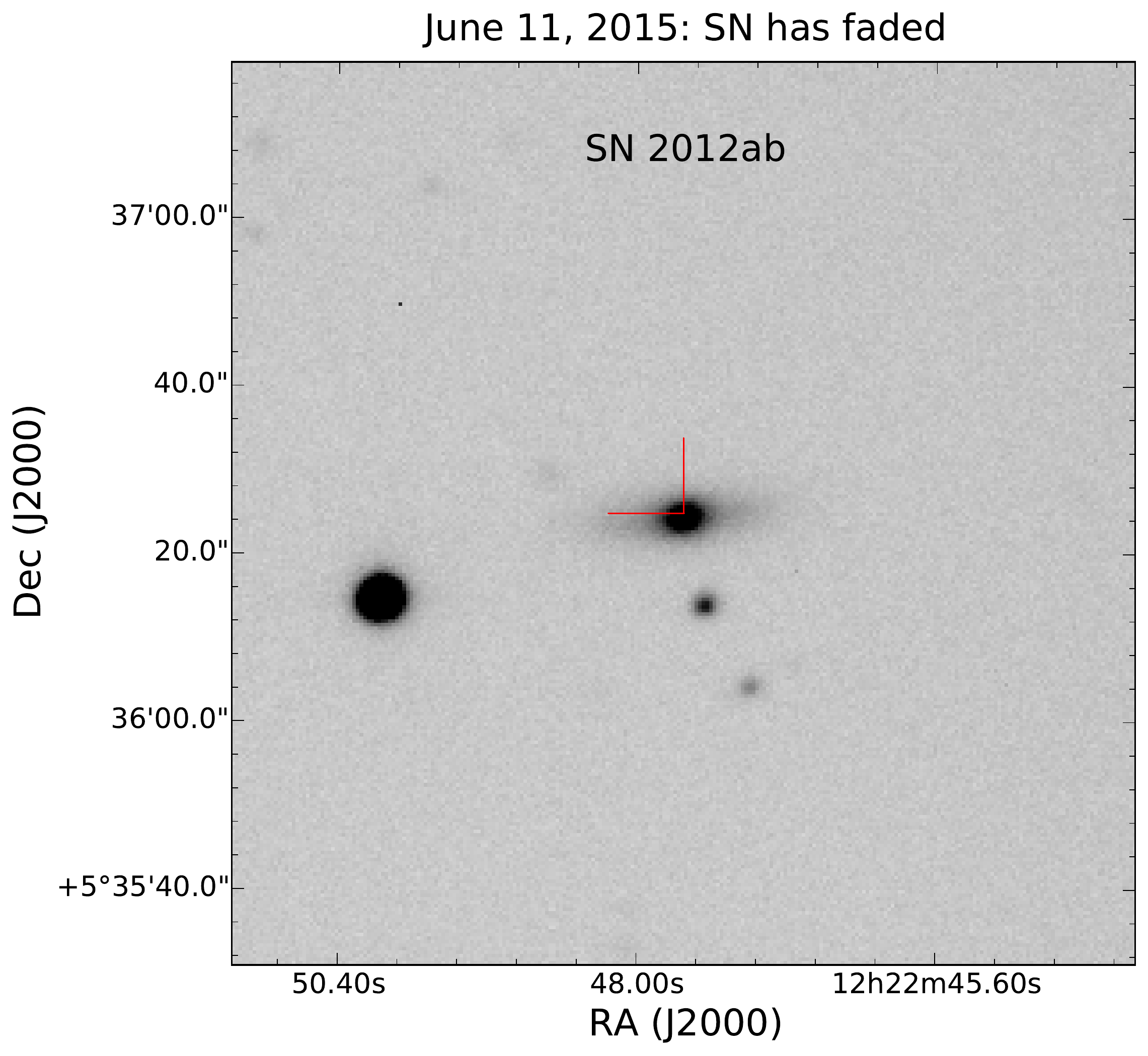}
\caption{{\it Left:} An unfiltered ROTSE-IIIb image of the host galaxy of SN~2012ab taken on 2012 February 3 with the SN present.  The hash marks show the location of the SN that we measure.  {\it Right:} An $R$-band image of the host galaxy taken with the Mont4K imager on the Kuiper telescope on 2015 June 11.  The SN had faded by the time of this image, but we mark its location near the nucleus.}
\label{fig:SN2012abfinder}
\end{figure*}

\begin{table}
\caption{Photometry of SN~2012ab}
\label{tab:ROTSELightCurve}
\begin{tabularx}{0.45\textwidth}{m{1.8cm}m{1.8cm}m{1.8cm}m{1.8cm}}
  \hline
MJD	&	$\mathrm{Day^{a}}$	&	Magnitude	&	$\sigma$\,(mag) \\
	\hline
55955.36	&	-1.99	&	(18.3)	&	    \\
55956.36	&	-0.99	&	(18.7)	&	    \\
55957.37	&	0.02	&	(17.4)	&	    \\
55958.36	&	1.01	&	16.13	&	0.05    \\
55959.36	&	2.01	&	15.57	&	0.03    \\
55960.34	&	2.99	&	15.39	&	0.05    \\
55971.41	&	14.06	&	15.44	&	0.03    \\
55972.35	&	15.00	&	15.73	&	0.03    \\
55973.31	&	15.96	&	15.53	&	0.05    \\
55976.31	&	18.96	&	15.53	&	0.03    \\
55978.30	&	20.95	&	15.38	&	0.03    \\
55979.30	&	21.95	&	15.53	&	0.02    \\
55982.29	&	24.94	&	15.52	&	0.03    \\
55984.28	&	26.93	&	15.32	&	0.03    \\
55986.35	&	29.00	&	15.40	&	0.06    \\
55987.27	&	29.92	&	15.49	&	0.04    \\
55988.29	&	30.94	&	15.53	&	0.05    \\
55990.26	&	32.91	&	15.71	&	0.05    \\
55992.25	&	34.90	&	15.36	&	0.12    \\
55993.25	&	35.90	&	15.54	&	0.10    \\
55997.29	&	39.94	&	15.94	&	0.06    \\
55998.24	&	40.89	&	15.33	&	0.04    \\
56002.24	&	44.89	&	15.82	&	0.04    \\
56009.22	&	51.87	&	15.39	&	0.05    \\
56010.22	&	52.87	&	15.78	&	0.03    \\
56012.28	&	54.93	&	15.63	&	0.18    \\
56014.20	&	56.85	&	16.06	&	0.04    \\
56016.20	&	58.85	&	15.92	&	0.07    \\
56017.20	&	59.85	&	16.29	&	0.08    \\
56019.19	&	61.84	&	15.76	&	0.12    \\
56021.18	&	63.83	&	16.02	&	0.08    \\
56028.26	&	70.91	&	16.40	&	0.13    \\
56034.23	&	76.88	&	16.54	&	0.11    \\
56035.26	&	77.91	&	17.03	&	0.12    \\
56041.16	&	83.81	&	(16.4)	&	   \\
56047.14	&	89.79	&	(16.5)	&	    \\
56048.22	&	90.87	&	(16.7)	&	    \\
56064.17	&	106.82	&	(16.5)	&	    \\
	\hline
	\end{tabularx}
	\tablecomments{\parbox{8cm}{\raggedright $^{a}$Days since discovery (2012 January 31.35 UT).\\  ROTSE unfiltered photometry calibrated as R-band.  Parentheses indicate upper limits.}}
\end{table}

\subsection{Spectroscopy}
\begin{figure*}
\centering
\includegraphics[width=1\textwidth,clip=true,trim=0cm 0cm 0cm 0cm]{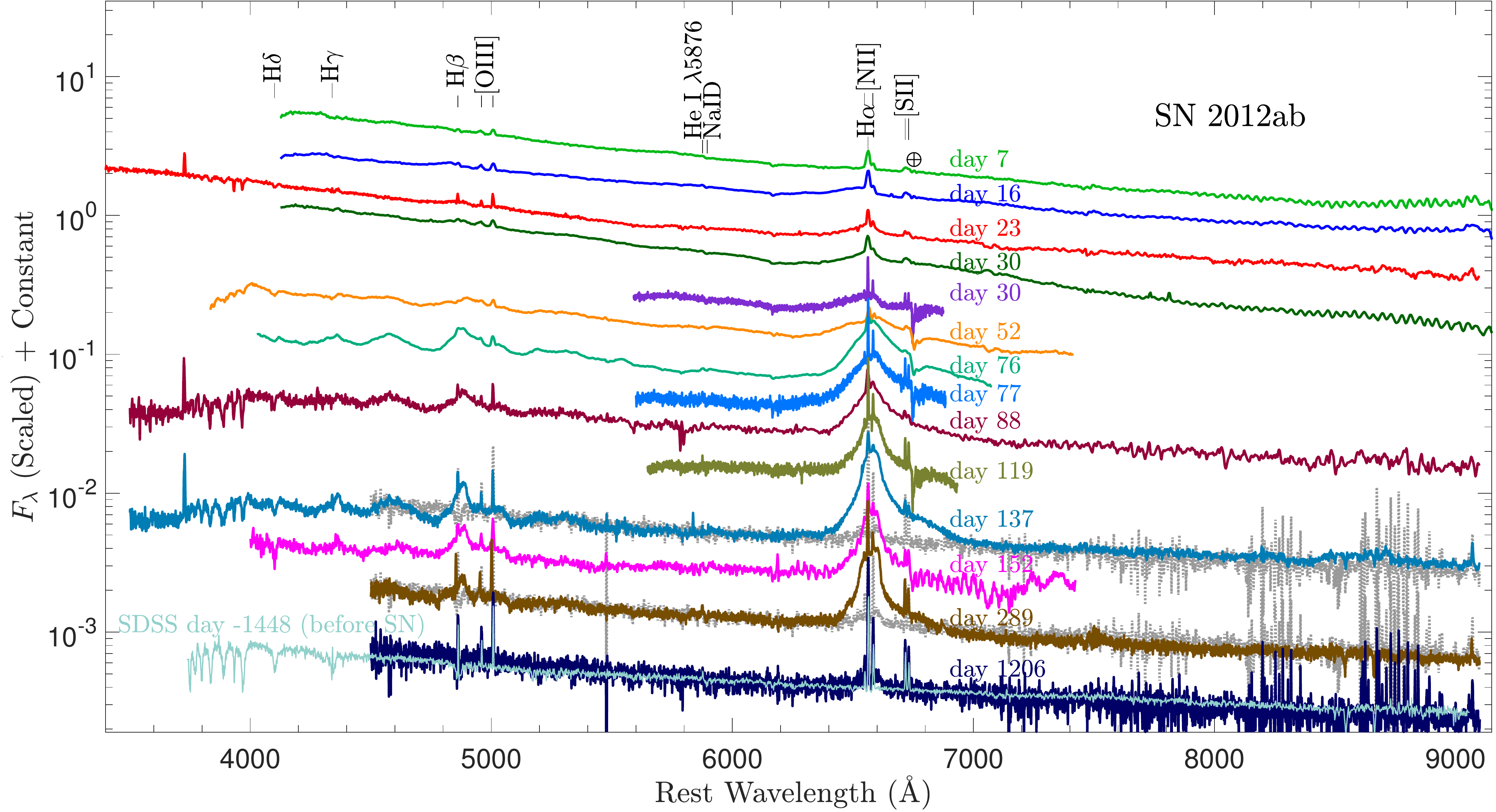}
\caption{Spectroscopic observations of SN~2012ab, dereddened ($E_{B-V} = 0.079$\,mag) and scaled for clarity (see Table \ref{tab:spectra}).  An SDSS spectrum taken 1448 days prior to the SN is overlaid on the very late time Keck spectrum (day 1206) to show that we are seeing primarily host galaxy emission at these late times \citep{2009ApJS..182..543A}.  We also plot a smoothed version of the day 1206 spectrum behind the day 137 and 289 spectra to show that the host galaxy nucleus accounts for most of the continuum emission in these spectra.   We find a continuum temperature in the pre-explosion SDSS spectrum of roughly $8{,}700$\,K $\pm\, 1{,}000$\,K based on blackbody fits to the dereddened blue continuum.}
\label{fig:allspectra}
\end{figure*}

We obtained optical spectra with a variety of instruments over the course of a year after the discovery of SN~2012ab.  Our spectra were taken on thirteen different nights with the Low-Resolution Spectrograph on the Hobby-Eberly Telescope \citep{LRS-HET}, the Kast spectrograph \citep{1993MillerStone} on the Shane 3-m reflector at Lick Observatory, the Bluechannel (BC) Spectrograph on the 6.5-m Multiple Mirror Telescope (MMT), the CCD Imaging/Spectropolarimeter (SPOL; \citealp{1992ApJ...398L..57S}) on the 2.3-m Bok Telescope and the MMT, the Deep Imaging Multi-Object Spectrograph \citep[DEIMOS;][]{2003SPIE.4841.1657F} on the 10-m Keck-II telescope, and the Low Resolution Imaging Spectrometer \citep[LRIS;][]{1995PASP..107..375O} at the 10-m Keck-I telescope.  The spectroscopic observations are detailed in Table \ref{tab:spectra} and the reduced spectra are shown in Figure \ref{fig:allspectra}.  

Atmospheric dispersion correctors were used with Keck-I/LRIS.  At all other instruments the data was taken either at low airmass or we oriented the slit along the parallactic angle \citep{1982PASP...94..715F} in order to minimize wavelength-dependent light losses and thus obtain correct relative spectrophotometry.

Standard spectral reduction procedures were followed for all of the spectra (see \S \ref{sec:Obs_SPOL} for SPOL polarization data details).  Since SN~2012ab was found near its host galaxy's nucleus, we took a very late-time spectrum on day 1206 in order to sample the likely contamination of host-galaxy light in all of the spectra.  We would have directly subtracted the host-galaxy light seen on day 1206 from all previous epochs of spectra, but the absolute flux calibration of the spectra is uncertain owing to slit losses, variable seeing and slit sizes, and in some cases cloud cover.  Consequently, we use the day 1206 spectrum as a qualitative template for the host-galaxy light.  We note that even the day 1206 spectrum may have some lingering intermediate-width H$\alpha$ emission, suggesting that late-time CSM interaction is still ongoing several years later.  We also show an SDSS spectrum ($R \approx 2000$) acquired 1448 days prior to discovery of SN~2012ab for comparison \citep{2009ApJS..182..543A}.

\begin{table*}
\begin{minipage}{200mm}
\caption{Spectroscopic Observations of SN~2012ab}
\label{tab:spectra}
\begin{tabular}{cccccccc}
  \hline
MJD & Year-Month-Day	&	Day\tablenotemark{a}	&	Telescope/Instrument & Wavelength Range ({\AA})\tablenotemark{b} & $\sim R\,(\lambda/\Delta\lambda)$\\
	\hline
55964	&	2012-02-07	&	7				&	HET/LRS	&							4,124--10,018 & 600\\
55973	&	2012-02-16	&	16			&	HET/LRS	&							4,126--10,020 & 600\\
55980	&	2012-02-23	&	23			&	Lick/Kast	&						3,385--10,077 & 600\\
55987	&	2012-03-01	&	30			&	HET/LRS	&							4,126--10,020 & 600\\
55987	&	2012-03-01	&	30			&	MMT/Blue Channel	&		5,591--6,874  & 3,300\\
56009	&	2012-03-23	&	52			&	Bok/SPOL	&						3,835--7,417  & 200\\
56033	&	2012-04-16	&	76			&	MMT/SPOL	&						4,029--7,075  & 200\\
56034	&	2012-04-17	&	77			&	MMT/Blue Channel	&		5,600--6,884  & 3,300\\
56045	&	2012-04-28	&	88			&	Lick/Kast	&						3,369--10,560 & 600\\
56076	&	2012-05-29	&	119			&	MMT/Blue Channel	&		5,649--6,933  & 3,300\\
56094	&	2012-06-16	&	137			&	Keck/LRIS	&						3,250--9,937  & 1,100\\
56109	&	2012-07-01	&	152			&	MMT/Blue Channel	&		3,777--8,916  & 700\\
56246	&	2012-11-15	&	289			&	Keck/DEIMOS	&					4,321--9,510  & 6,000\\
57163	&	2015-05-21	&	1206		&	Keck/DEIMOS	&					4,326--9,468  & 6,000\\
	\hline
	\end{tabular}
	\tablenotetext{a}{Days since discovery (2012 January 31.35 UT).}
	\tablenotetext{b}{After applying host-galaxy redshift.}
\end{minipage}
\end{table*}

\subsection{Spectropolarimetry}
\label{sec:Obs_SPOL}
We obtained spectropolarimetric observations of SN~2012ab using the CCD Imaging/Spectropolarimeter (SPOL; \citealp{1992ApJ...398L..57S}) on the 90-inch Bok (2012 Mar. 23) and 6.5-m MMT telescopes (2012 Apr. 16).  We used a 3.0\arcsec\ slit at the Bok and a 1.9\arcsec\ slit at the MMT.  Because SN~2012ab was located in the nucleus of its host galaxy, using such large slit widths means that we have significant contamination from galaxy light in all of our spectra.  This contamination is particularly important at late times when the continuum light from CSM interaction has declined significantly.  We used the 600\,lines\,mm$^{-1}$ grating, which has a typical wavelength coverage of $\sim 3900$--7550\,{\AA}  with a resolution of $\sim 20$\,{\AA} ($\sim 900$\,km\,s$^{-1}$).  A rotatable semi-achromatic half-wave plate modulates the incident polarization and a Wollaston prism in the collimated beam separates the orthogonally polarized spectra onto a thinned, antireflection-coated $800\times1{,}200$ pixel SITe CCD.  SPOL has a fully-polarizing Nicol prism in the beam above the slit which corrects for the efficiency of the waveplate as a functon of wavelength.  A series of four separate exposures that sample 16 orientations of the waveplate yield two independent, background-subtracted measures of each of the normalised linear Stokes parameters, $Q$ and $U$.  We acquired two such sequences at the Bok and three at the MMT.  We then combined each set of sequences by epoch to improve the signal-to-noise ratio (S/N).

We used Hiltner 960 and VI~Cyg~12 as polarimetric standards \citep{1992AJ....104.1563S} to confirm that the instrumental polarization for SPOL at the Bok and MMT telescopes was $<0.1\%$ for each epoch.  We also measured the polarization angle, $\theta$, of these polarimetric standards in order to calibrate our data values to the standard equatorial frame.  The discrepancy between the measured and the expected position angle was $<0.2\degree$ for each of the polarimetric standard stars.

\section{Results}
\label{sec:Res}

\subsection{Extinction and Reddening}
\label{sec:Ext}

\begin{figure}
\centering
\includegraphics[width=0.48\textwidth,clip=true,trim=0cm 0cm 0cm 0cm]{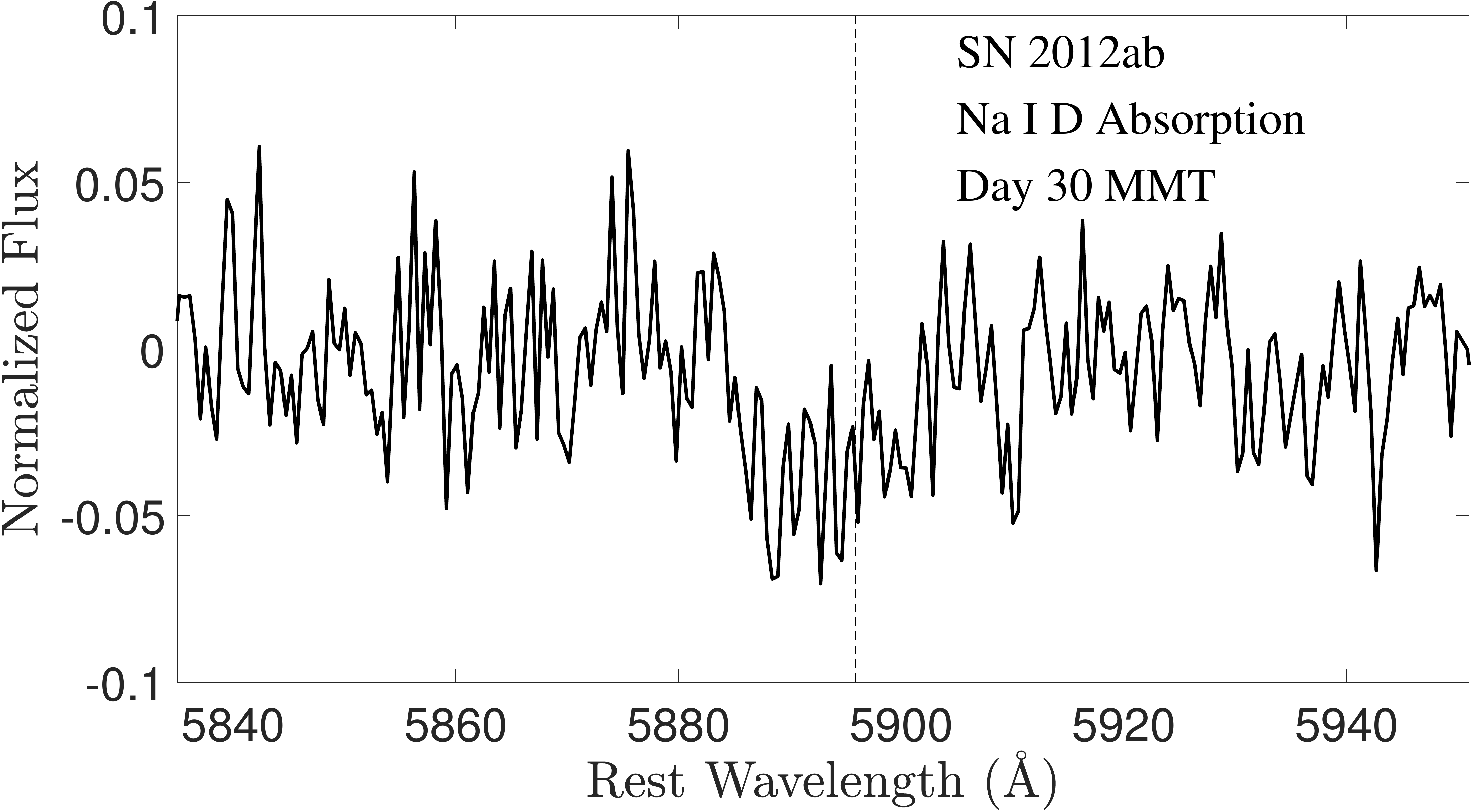}
\caption{A moderately high-resolution spectrum taken on day 30 showing the sodium doublet absorption which we use to constrain the host-galaxy extinction and interstellar polarization.  Vertical dashed lines show the rest wavelengths of the Na \textsc{i} D lines.}
\label{fig:NaID}
\end{figure}

The strength of the absorption lines of Na \textsc{i} D $\lambda \lambda 5890 (D1), 5896 (D2)$ correlates with the interstellar dust extinction present along a particular line of sight.  While this relation does not perform well with low-resolution spectra \citep{2011MNRAS.415L..81P}, it can be used with moderate-resolution spectra when the Na \textsc{i} D2 line is not saturated and the doublet is not blended \citep{2012MNRAS.426.1465P}.  \citet{2013ApJ...779...38P} found that the sodium doublet absorption for one-fourth of their sample of SNe Ia was stronger than expected for dust extinction values estimated from SN color.  In our moderately high-resolution spectrum on day 30, the sodium doublet is not blended together and we measure the equivalent widths for the D1 line ($\lambda 5896$) and the D2 line ($\lambda 5890$) to be $0.25 \pm 0.04$ \r{A} and $0.29 \pm 0.04$ \r{A}, respectively (see Figure \ref{fig:NaID}.  Based on these equivalent widths, the relation provided by \citet{2012MNRAS.426.1465P} suggests that we have an additional extinction along the line of sight caused by the host galaxy of $A_V = 0.19$\,mag (assuming $A_V = 3.08 E_{B-V}$; \citealp{1992ApJ...395..130P}) or $A_R = 0.15$\,mag.  Many other attempts have been made in the literature to connect the equivalent width of the absorption in the sodium doublet to extinction \citep{1994AJ....107.1022R,1997A&A...318..269M,2003fthp.conf..200T,2012MNRAS.426.1465P}, but our results are not significantly changed even if we choose the the model with the highest estimated extinction of $A_V = 0.44$\,mag instead.  We adopt a total Milky Way \citep{2011ApJ...737..103S} plus host-galaxy extinction of $A_V = 0.24$\,mag ($E_{B-V}=0.079$\,mag) or $A_R = 0.20$\,mag.
 
Figure \ref{fig:allspectra} shows spectra dereddened by $E_{B-V}=0.079$\,mag.  After this correction, the day 23 spectrum exhibits a continuum temperature of $\sim 12{,}000$\,K $\pm 3{,}000$\,K, similar to that of other SNe~IIn at early times.

\subsection{Light Curve}
The unfiltered light curve ($\sim$ \textit{R}-band) for SN~2012ab obtained by ROTSE-IIIb is displayed in Figure \ref{fig:ROTSELightCurve}.  The absolute magnitudes shown have been adjusted for Milky Way and host-galaxy reddening (determined from Na \textsc{i} D line widths as discussed in \S \ref{sec:Ext}) and for the distance modulus of $m-M = 34.57$\,mag based on a distance to the host galaxy of $D=82.3\pm 5.8\,\mathrm{Mpc}$.   

Upper limits on SN~2012ab set $\sim 1$ day prior to discovery show that the rise to peak for this object must have been very fast.  However, our upper limits are not very deep and our lack of detection is still consistent with a normal or faint Type II-P SN.  After being discovered at an absolute magnitude of $-18.7$, SN~2012ab reached a peak absolute magnitude of $-19.4$ within only 3 days of discovery.  This quick rise suggests that SN~2012ab may have exploded prior to our detection and we are just now seeing the CSM interaction causing this rapid brightening, as in the case of SN~2009ip when the SN was initially faint \citep{2013MNRAS.430.1801M,2013ApJ...763L..27P,2013ApJ...767....1P}.

Although the photometric points exhibit large scatter, the light curve appears relatively constant at above $-19.5$\,mag until about day 55.  Beyond day 50 the light curve begins to decline more rapidly ($\sim 0.06$\,mag $\mathrm{day^{-1}}$  between days 52 and 78), but we do not have very deep or late-time constraints that allow us to accurately track this decline into the nebular phase of the SN.  Our spectra show that the SN has clearly faded, but accurate spectrophotometry is not possible because of slit losses, galaxy nucleus contamination, and variable seeing.

\subsection{SN Location}
\label{sec:Res:SNLoc}
We find astrometric fits to both our ROTSE and Kuiper images (Figure \ref{fig:SN2012abfinder}) using \url{astrometry.net} \citep{2008AJ....135..414B}.  We then measure the location of the SN from the ROTSE image and the location of the host galaxy from the Kuiper image using radial profile fits to a Moffat distribution.  We determine the uncertainty in the location of the centroid by replicating the noise level in each image and refitting the centroid 100 times.  The location of the SN is measured to be $\alpha\mathrm{(J2000)} = 12^\mathrm{h}22^\mathrm{m}47^\mathrm{s}.63$, $\delta\mathrm{(J2000)} = +05\degree36^{\prime}24^{\prime\prime}.83 \pm 0.23^{\prime\prime}$, and the location of the host galaxy is measured to be $\alpha\mathrm{(J2000)} = 12^\mathrm{h}22^\mathrm{m}47^\mathrm{s}.64$, $\delta\mathrm{(J2000)} = +05\degree36^{\prime}24^{\prime\prime}.41 \pm 0.02^{\prime\prime}$.  The difference between these is $\Delta\alpha = 0.01^\mathrm{s} \pm 0.22^\mathrm{s}$ and $\Delta\delta = 0.42^{\prime\prime} \pm 0.078^{\prime\prime}$.  $\Delta\delta$ is bigger than the uncertainty in the $\Delta\delta$.  This suggests that the SN is not coincident with the host-galaxy nucleus as initially reported, but rather is offset by roughly $0.42^{\prime\prime}$.  At a distance of 82.3\,Mpc, the SN has a projected distance from the nucleus of the host galaxy of $\sim 160$\,pc.

\subsection{Spectral Morphology}
We see a strong blue continuum ($\sim 12{,}000$\,K $\pm 3{,}000$\,K) present at early times (day 30) in our spectra.  Both the pre-discovery SDSS spectrum and the very late-time Keck spectrum on day 1206 show a strong blue continuum, though at a cooler temperature of $\sim 9000$\,K $\pm 3000$\,K.  Very late-time photometry of the galaxy nucleus taken on days 1176 and 1228 reveals an $m_R \approx 17.7$\,mag source, which we assume is primarily light from the host-galaxy nucleus.  This host-galaxy emission accounts for roughly $10\%$ of the total light from the SN at peak magnitude and about half of the total light in our spectra around day 80.  We do not estimate the SN magnitudes at late times from our spectra because of uncertainty in slit losses and other factors mentioned above, most of which make the absolute flux calibration difficult when the source is near the nucleus of its host.

We detect a number of spectral features in the fourteen different spectra.  Most prominent are the resolved H$\alpha$ and H$\beta$ emission lines present from day 7 to day 1{,}206.  We also detect several narrow nebular lines in many of the spectra: these include [O \textsc{iii}] $\lambda\lambda$4959, 5007, [N \textsc{ii}] $\lambda\lambda$6548, 6583, and [S \textsc{ii}] $\lambda\lambda$6717, 6731.  

\begin{figure}
\centering
\includegraphics[width=0.48\textwidth,clip=true,trim=0cm 0cm 0cm 0cm]{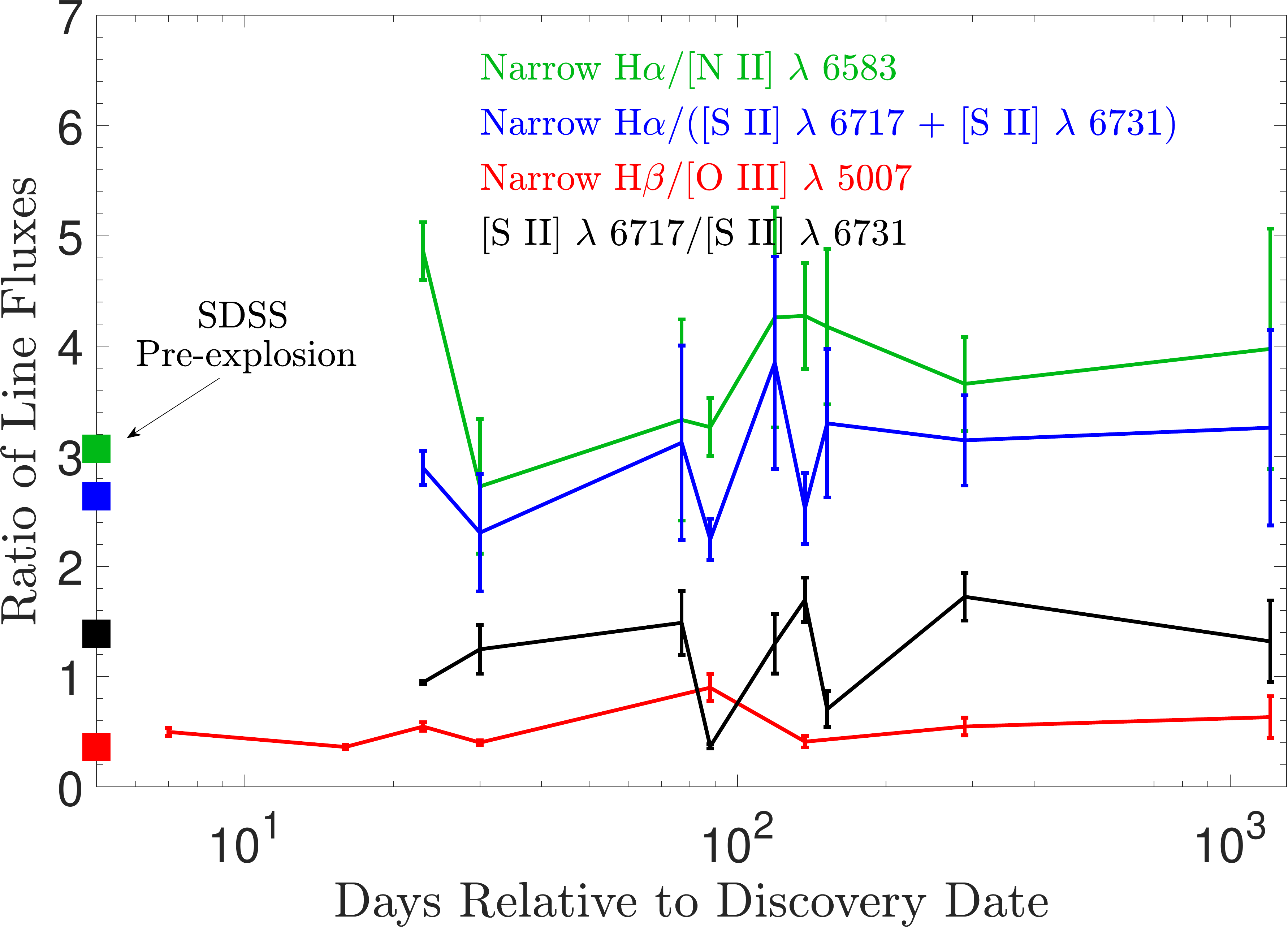}
\caption{A comparison of various nebular line-intensity ratios with time.  The filled squares indicate the line ratios for the SDSS spectrum 1448 days prior to discovery.  The ratio of the narrow H$\alpha$ and H$\beta$ lines to nebular lines does not change significantly over the course of most of the evolution of SN~2012ab.}
\label{fig:Nebular}
\end{figure}

In order to determine if the narrow H$\alpha$, which has a full width at half-maximum intensity (FWHM) of less than 300\,km\,$\mathrm{s^{-1}}$, is being emitted by a slowly moving wind or by a distant nebular region, we take the ratios of H$\beta$ to [O \textsc{iii}] $\lambda 5007$, H$\alpha$ to [N \textsc{ii}] $\lambda 6583$, and H$\alpha$ to [S \textsc{ii}] $\lambda\lambda$6717, 6731 (Figure \ref{fig:Nebular}).  Except perhaps for the day 23 spectrum, the ratio of the narrow Balmer flux to the nebular lines does not change significantly (though the error bars are large on all of these measurements).  We unfortunately exclude the earliest spectra we have from this comparison because the nebular lines are heavily blended with the Balmer lines and, as a result, we cannot reliably measure their individual flux levels.  The constant line ratios persisting as the SN expands and fades suggest that most of the narrow H$\alpha$ and H$\beta$ emission may be nebular and not dense CSM being overtaken by the shock.  This is further supported by the fact that both the pre-SN SDSS spectrum obtained 1448 days prior to discovery and the late-time spectrum dominated by the host galaxy on day 1206 show these nebular lines with similar ratios (Figure \ref{fig:Nebular}).  Based on the ratios of the nebular lines, a Baldwin, Phillips, \& Terlevich (BPT) diagram indicates that the host is consistent with an H\,\textsc{ii} region, not a LINER or narrow-line active galactic nucleus (AGN).

H$\alpha$ and H$\beta$ have broad (FWHM $\approx$ 20${,}$000\,km\,$\mathrm{s^{-1}}$) and intermediate-width (FWHM of $\approx$ 4${,}$500\,km\,$\mathrm{s^{-1}}$) emission components which shift from the blue side at early epochs to the red side at later epochs.  H$\alpha$ shows no broad absorption features in any of our spectra.  H$\beta$ seems to have some broad absorption features on days 76-137, though this may be a trough between two emission features rather than an actual absorption feature.  There are other broad features seen in the spectra, but they are significantly weaker than those in a normal SN photosphere.  The evolution of the broad H$\alpha$ and H$\beta$ lines is discussed later in \S \ref{sec:Dis:asymHalpha}.

\subsection{Spectropolarimetry}
\label{sec:Res:Specpol}
Our spectropolarimetric analysis is performed primarily using the linear Stokes parameters, $q=Q/I$ and $u=U/I$, which are rotated $45\degree$ with respect to each other, allowing us to decompose the polarization signal into orthogonal components.  Typically, one can combine the Stokes parameters to obtain the polarization level, $P = \sqrt{Q^2 + U^2}$, and the position angle on the sky, $\theta = 1/2\, \mathrm{tan}^{-1} (U/Q)$.  However, since the definition of the polarization angle makes it a positive-definite value, it is biased in cases where we have low S/N because fluctuations will raise the mean polarization level significantly.  In order to attempt to control for this effect, we use a debiased polarization level, $P_{db} = \sqrt{|Q^2 + U^2 - \frac{1}{2}(\sigma_Q^2 + \sigma_U^2)|}$ \citep{1974ApJ...194..249W}.  Even the debiased polarization level is not a perfect measure \citep{1988MillerSPOL}, so we attempt to perform our analysis in the $q$ and $u$ plane whenever possible.  This allows us to avoid problems with the positive-definite nature of polarization and see where the changes in polarization signal are most pronounced.

We must address the tricky issue of interstellar polarization (ISP) in order to determine the polarization intrinsic to SN~2012ab.  It is difficult to measure the combined level of the ISP arising from the Milky Way and from the SN host galaxy.  Fortunately however, a variety of circumstances point to a very low ISP for SN~2012ab.

First, we see heavy depolarization of the H$\alpha$ and H$\beta$ lines on day 76, in contrast to the polarization seen in the H$\alpha$ and H$\beta$ lines on day 52.  The position angle on the sky does not change across the H$\alpha$ and H$\beta$ lines in the MMT spectropolarimetry, suggesting that the depolarization is caused by a dilution of the polarized continuum with unpolarized line emission.  If unscattered line emission dominates the light near line peak, the polarization signal should approach the ISP \citep{2008ApJ...688.1186H,2011A&A...527L...6P}.  The peak of the H$\alpha$ line suggests an ISP value of $< 0.5\%$.

In \S \ref{sec:Ext} we found a low value for the reddening of $E_{B-V} = 0.079$\,mag based on Na \textsc{i} D absorption plus Milky Way reddening.  We adopt this total extinction level when dereddening our spectra in Figure \ref{fig:allspectra} and all subsequent analysis.  Additionally, the Serkowski relation suggests that ISP $ < 9E(B-V)$ for Milky Way dust \citep{1975ApJ...196..261S}, which means that we can use the measure of $E(B-V)$ from the Na \textsc{i} D absorption lines and the Milky Way to place a constraint on the level of the ISP to $<0.71\%$.  This constraint is roughly consistent with the observed depolarization across the narrow H$\alpha$ line.

\begin{figure*}
\centering
\includegraphics[width=1\textwidth,clip=true,trim=0cm 0cm 0cm 0cm]{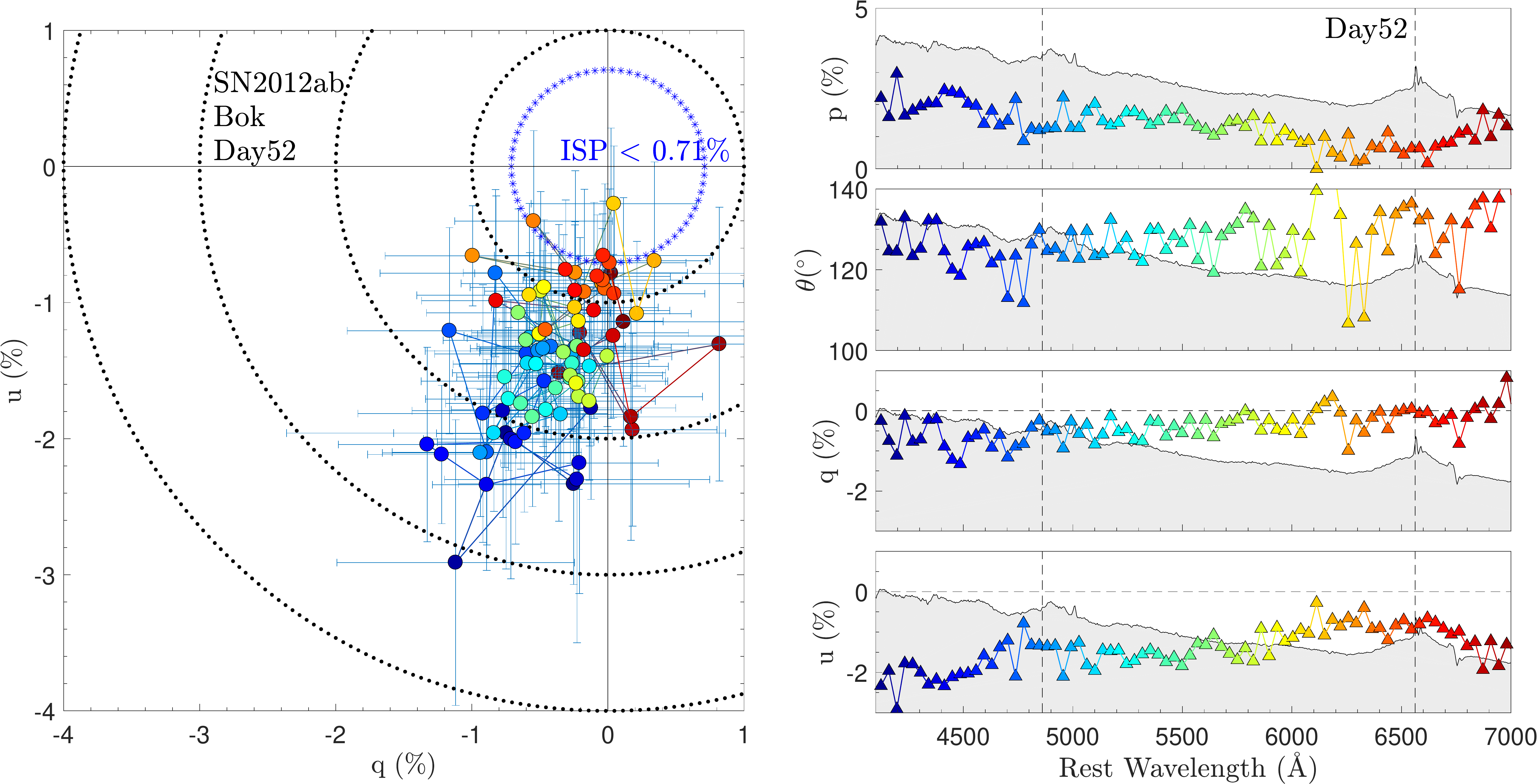}
\includegraphics[width=1\textwidth,clip=true,trim=0cm 0cm 0cm 0cm]{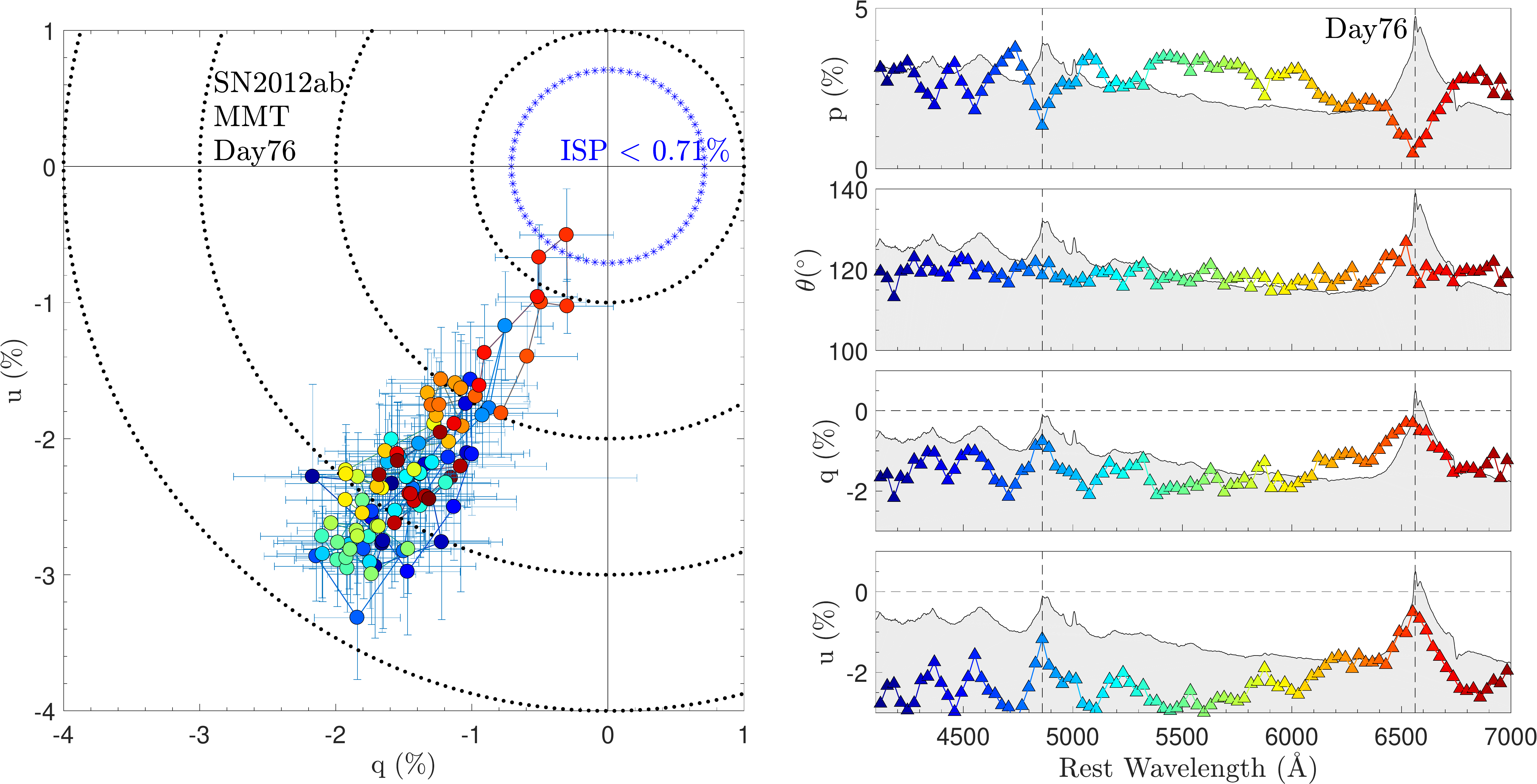}
\caption{{\it Top panels:} $q$--$u$ Stokes parameters, polarization $p$, and position angle $\theta$ for SN~2012ab from the 90-inch Bok telescope on day 52.  The data are grouped into $\sim 30$\,{\AA} bins.  Shaded regions show a scaled version of the flux spectrum.  We have adopted an ISP value of $<0.71\%$ based on Na \textsc{i} D absorption-line measurements (see \S \ref{sec:Obs_SPOL}).  {\it Bottom panels:} The same for the 6.5-m MMT data on day 76.  Colours in the $q$--$u$ plots correspond to wavelengths as shown on the right.}
\label{fig:QUcircle}
\end{figure*}

Figure \ref{fig:QUcircle} shows the spectropolarimetric data plotted in the $q$--$u$ plane.  Both the day 52 and day 76 spectra exhibit a relatively strong level of continuum polarization.  We measure the continuum polarization in two regions (5{,}400--5{,}500\,{\AA} and 6{,}100--6{,}200\,{\AA}; see \citealt{2008ApJ...688.1186H}) where the spectrum is devoid of line emission.  On day 52, we measure the continuum polarization to be $1.7\%\pm0.1\%$ at 5{,}400--5{,}500\,{\AA} and  $0.8\%\pm0.1\%$ at 6{,}100--6{,}200\,{\AA}.  By day 76, the continuum polarization has risen to $3.5\%\pm0.1\%$ at 5{,}400--5{,}500\,{\AA} and $2.3\%\pm0.1\%$ at 6{,}100--6{,}200\,{\AA}.  There is a slight change in the position angle (across the entire range 4{,}100--7{,}000\,{\AA}) from $127^{\circ}\pm15^{\circ}$ on day 52 to $119^{\circ}\pm9^{\circ}$ on day 76, though the change is not sufficiently large to indicate a statistically significant drop.

\section{Discussion}

\subsection{SN, AGN, or TDE?}
Although we find that SN~2012ab is offset from its host galaxy by roughly $0.4^{\prime\prime}$ (see \S \ref{sec:Res:SNLoc}), we also consider the possibility that it is an AGN or TDE based on the properties of the transient itself.  We find it unlikely that SN~2012ab is an AGN by considering both the light curve and various emission-line strengths.  As we discuss in \S \ref{sec:Dis:lc}, SN~2012ab has a relatively smooth light curve, but it shows a significant drop of over 1.5\,mag between days 52 and 78.  AGN can exhibit large variations in magnitude, but generally take months or years to show this great a change \citep[e.g.,][]{1997ARA&A..35..445U}.  On the timescale of days to weeks, the variability of AGN in the optical is usually much less pronounced than we see in SN~2012ab \citep{1997ARA&A..35..445U}.  Our predetection upper limits  imply a steep rise in the light curve.  This is followed by $\sim 50$ days of nearly constant magnitude and then a significant and steady decline in the light curve.  These characteristics are dissimilar to variability often seen in AGN \citep{2001sac..conf....3P}.  Also, at the time of writing, SN~2012ab has not rebrightened.  

While we see a variety of nebular lines that are likely from a nearby H~II region in our spectra, we would expect to see a greater range of ionization levels in an AGN.  These forbidden lines include [O~I] $\lambda\lambda$6300, 6364, [N~I] $\lambda$5199, and [Fe~VII] $\lambda$5721, none of which we detect.  The He~II $\lambda$4686 line in AGN is often quite strong even compared to the Balmer series.  We do not detect any He~II $\lambda$4686 in our spectra.  Lastly, the pre-SN SDSS spectrum shows no broad component of H$\alpha$ and the very late-time Keck spectrum shows a weak broad component of H$\alpha$.  All of our spectra in between these two, however, contain a strong broad H$\alpha$ emission component.  While some AGN show a broad line region and others do not, it is not typical for AGN to transition from one to the other and back again, further suggesting that this galaxy is not an AGN.

While TDEs are not yet as well studied as AGN, we do expect that they would show evidence of asphericities \citep{2009MNRAS.400.2070S} like those we see in SN~2012ab (see \S 4.4).  \citet{2009MNRAS.400.2070S} predict the optical signatures of TDEs to have peak luminosities of $\sim 10^{43}$--$10^{44}$\,$\mathrm{erg s^{-1}}$ and characteristic decay timescales of $\sim 10$ days.  Although these luminosities are comparable to that of SN~2012ab, the theoretical decline rate in TDEs ($t^{-5/3}$; \citealp{1988Natur.333..523R,1989ApJ...346L..13E,1989IAUS..136..543P}; shown in Figure \ref{fig:ROTSELightCurve}) is inconsistent with our light curve of SN~2012ab.  When we match the theoretical TDE decline to the measured decline of SN~2012ab, we find that a TDE event would require unreasonably bright early-time magnitudes.  For instance, we would have expected the source to be brighter than $-20$ mag for all days prior to day 47.  Even when considering the optical decline rates observed by \citet{2011MNRAS.410..359L} for TDEs just after peak ($t^{-2.6}$) or a few months after peak ($t^{-5/12}$), we still do not see similar behaviour in the light curve of SN~2012ab.  

A lack of forbidden emission lines in TDEs is expected owing to very high densities in the emitting regions \citep{2009MNRAS.400.2070S}.  This is consistent with our narrow emission lines arising from nearby H~II regions.  The spectrum of a TDE should consist of broad emission lines offset in redshift from the host galaxy's emission lines \citep{2009MNRAS.400.2070S}. The spectra of SN~2012ab do show some offset between the narrow lines and the intermediate-width and broad lines.  \citet{2012Natur.485..217G} suggest that we would see strong He II emission in TDEs while \citet{2014ApJ...793...38A} claim that these emission lines would be likely.  \citet{2012Natur.485..217G} also provide various X-ray and ultraviolet criteria that help distinguish SNe, AGN, and TDEs, but we do not have any such data to aid in our classification process for SN~2012ab.  Overall, our data are most consistent with classifying SN~2012ab as a core-collapse SN, though we cannot rule out the possibility of SN~2012ab being a highly unusual TDE.  The observed positional offset from the nucleus, however, makes a TDE or AGN hypothesis even less likely.

\subsection{Light Curve}
\label{sec:Dis:lc}
Having peaked at an absolute magnitude of $M=-19.5$ and persisted at brighter than $M<-18$\,mag for over 75 days, SN~2012ab was most likely a luminous Type II SN.  At peak, it is over a magnitude brighter than a standard SN~II-L (SN~2003hd) and $\sim 3$\,mag brighter than a standard SN~II-P (SN~1999em), but about the same as SN~1998S (Type IIn).  SNe~II-L are generally seen to have decline rates of $>0.01$\,mag\,$\mathrm{day^{-1}}$ in the $I$ \citep{2009ApJ...694.1067P} and $V$ \citep{2014MNRAS.445..554F} bands, while SNe~II-P during the 100\,day plateau decline more slowly.  As we discuss in greater detail in \S \ref{sec:Dis:compareSNeII}, whether SNe~II-P and II-L should be considered distinct classes is still a topic of debate \citep{2012ApJ...756L..30A,2014ApJ...786...67A,2016MNRAS.459.3939V}.  For the first $\sim50$ days, the light curve of SN~2012ab is essentially constant.  Between days 52 and 78, the light curve drops at an average rate of $\sim 0.06$\,mag $\mathrm{day^{-1}}$.  

We estimate the total radiated energy over the course of our photometric coverage to be $E_{\rm rad} \approx 10^{50}\,\mathrm{erg}$.  In comparison, the average core-collapse SN has a total radiated energy of $E_{\rm rad} \approx 10^{48}$--$10^{49}\,\mathrm{erg}$.  The additional energy we see in our estimate is likely caused by CSM interaction which powers a significant fraction of the light curve for an extended period of time.

\subsection{CSM Interaction Luminosity}

The intermediate-width H$\alpha$ component seen in our spectra suggests the presence of CSM interaction in SN~2012ab.  If the CSM interaction is powering a large component of the luminosity, then we can estimate a lower limit to the wind-density parameter ($w = \dot{M}_{\rm CSM}/v_{\rm pre}$, where $v_{\rm pre}$ is the velocity of the preshock wind) prior to explosion for SN~2012ab. We calculate the wind-density parameter as 

\begin{equation}
w = 2L/v_{\rm post}^3,
\end{equation}

\noindent where $L$ is the observed luminosity and $v_{\rm post}$ is the velocity of the postshock shell \citep{2008ApJ...686..467S}.  In determining all of our input parameters, we will make conservative estimates that err on the side of a lower resulting wind-density parameter (or equivalently a lower derived mass-loss rate).  

We find the velocity of the postshock shell by looking at the intermediate-width component of the H$\alpha$ line profile which becomes prominent by day 76, but is more distinct from the broad component by day 137.  The intermediate-width component is measured to have a half width at half-maximum intensity of $2{,}200$\,km\,$\mathrm{s^{-1}}$ on the blue side and $3{,}300$\,km\,$\mathrm{s^{-1}}$ on the red side on day 77.  In order to provide a lower bound on $w$, we use $v_{\rm post}=3{,}300$\,km\,$\mathrm{s^{-1}}$.  We adopt a magnitude of $M_{R} = -18.24$ on day 77 based on interpolation between the actual dates we have photometry to estimate the luminosity ($L \approx 1.6 \times 10^{9}\,{\rm L}_{\odot}$).  We assume no bolometric correction and obtain a conservative estimate of the wind-density parameter of $9 \times 10^{17}$\,g\,$\mathrm{cm^{-1}}$.  If we assume a steady wind velocity from the progenitor of $v_{\rm pre} = 100$\,km\,$\mathrm{s^{-1}}$, we can estimate the mass-loss rate to be at least $\dot{M} = 0.050\,{\rm M}_{\odot}\,\mathrm{yr^{-1}}$.  As discussed in \S \ref{sec:Dis:asymHalpha}, the pre-SN wind seems to have reached a distance of $\sim 1{,}600$\,au on the far side prior to interaction with the SN ejecta, suggesting that mass loss was occurring $\sim 75$\,yr (depending heavily on the assumed $v_{\rm pre}$) prior to explosion.  

There are a number of suggested types of progenitor stars of SNe~IIn which have widely differing mass-loss rates.  \citet{2014ARA&A..52..487S} discussed progenitors of SNe~IIn such as luminous blue variables (LBVs) with mass-loss rates as high as 0.01--10\,M$_{\odot}\,\mathrm{yr^{-1}}$ or red supergiants and yellow hypergiants with mass-loss rates in the range of $10^{-4}$--$10^{-3}$\,M$_{\odot}\,\mathrm{yr^{-1}}$.  Of these, only LBVs in eruption are known to achieve mass-loss rates in the range of SN~2012ab.

\subsection{Asymmetric H$\alpha$}
\label{sec:Dis:asymHalpha}

\begin{figure}
\centering
\includegraphics[width=0.485\textwidth,clip=true,trim=0cm 0cm 0cm 0cm]{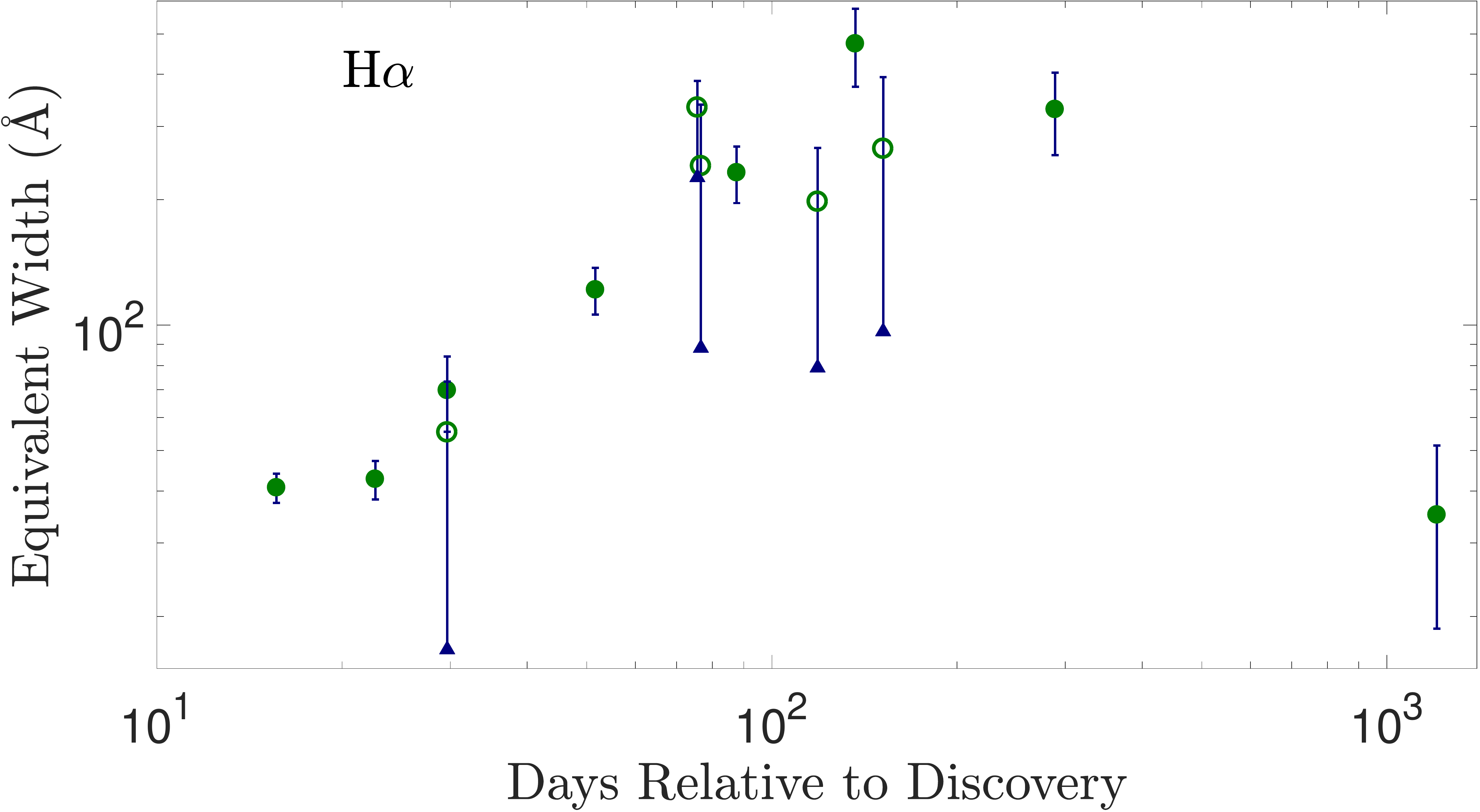}
\caption{H$\alpha$ broad-line equivalent width measurements.  Unfilled circles are estimates of the equivalent width from spectra that do not extend far into the red, making it hard to sample the continuum on the red side of the H$\alpha$ line.  Blue triangles indicate lower limits that are set by assuming that the flux level at the edge of the intermediate-width H$\alpha$ line on the red side is the continuum level on the red side.  See \S \ref{sec:Dis:asymHalpha} for a detailed discussion of the evolution of the H$\alpha$ line profile.}
\label{fig:HalphaEWFluxhigh}
\end{figure}

\begin{figure*}
\centering
\includegraphics[width=1.0\textwidth,clip=true,trim=0cm 0cm 0cm 0cm]{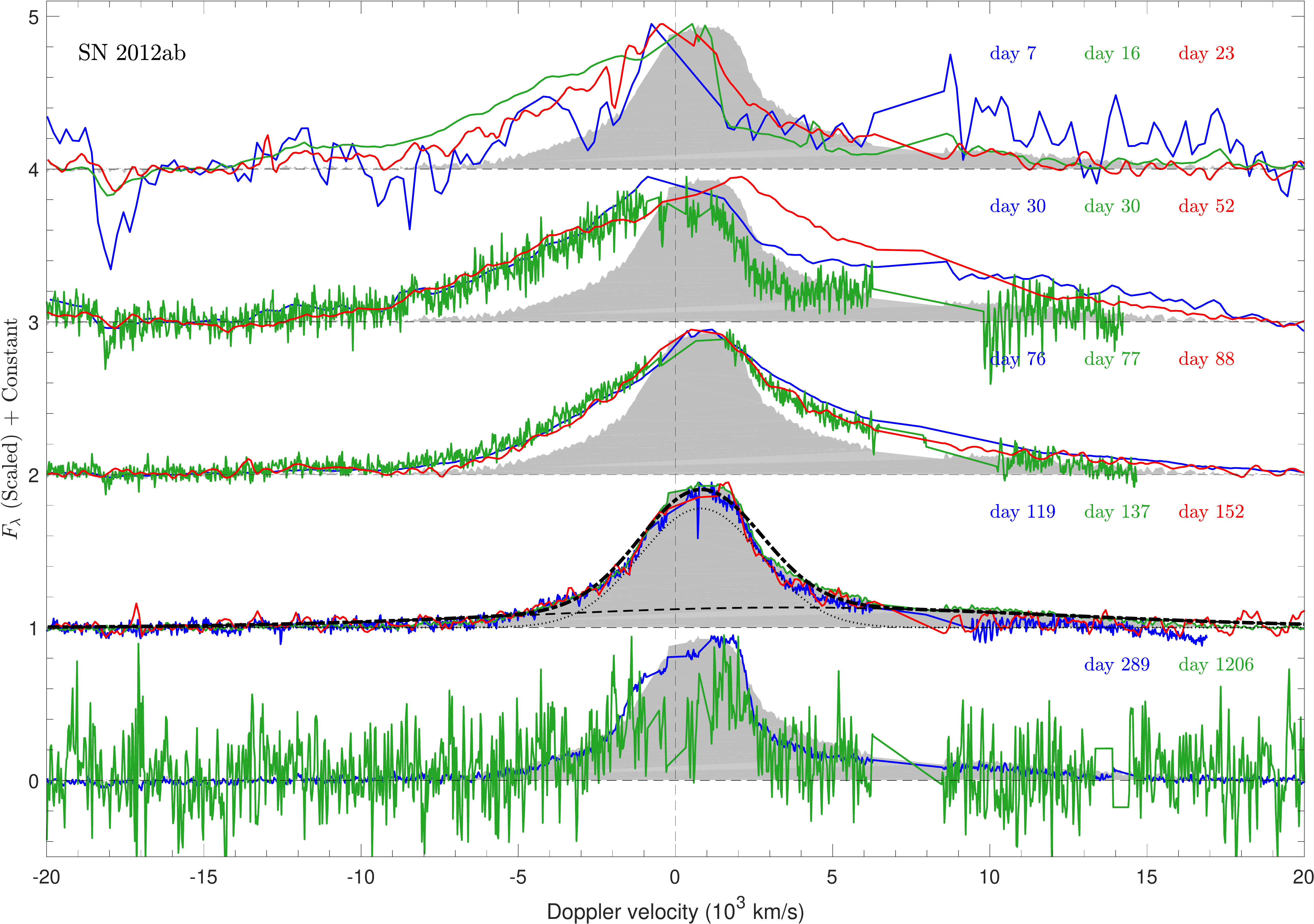}
\caption{Evolution of the intermediate-width and broad H$\alpha$ line profiles.  We have interpolated the data across nebular lines and the narrow H$\alpha$ component.  The peak of each profile was normalised and each triplet group of data was offset for clarity.  We constructed a template from the day 137 spectrum and plotted it as a shaded background in each panel.  The fits to the intermediate-width component (dotted), the broad component (dashed), and the sum of the two components (dash-dotted) are plotted in black.}
\label{fig:HalphaTripleCompare}
\end{figure*}

SN~2012ab exhibits strong broad and intermediate-width H$\alpha$ emission at all epochs up to and including day 289.  We find the best-fit Gaussian profiles to the intermediate-width and broad components of the day 137 spectrum simultaneously; see Figure \ref{fig:HalphaTripleCompare}.  The intermediate-width component is centred at 800\,km\,$\mathrm{s^{-1}}$ with a FWHM of 4${,}$500\,km\,$\mathrm{s^{-1}}$, and the broad component is centred at 2${,}$800\,km\,$\mathrm{s^{-1}}$ with a FWHM of 20${,}$000\,km\,$\mathrm{s^{-1}}$.  We measure the equivalent width of the broad H$\alpha$ emission in all of our spectra by removing the narrow component of the line (via interpolation), removing the continuum flux via a polynomial fit across the edges of the H$\alpha$ line, and then summing the remaining normalized flux in the H$\alpha$ line; our results are shown in Figure \ref{fig:HalphaEWFluxhigh}.  The unfilled circles represent estimates for the equivalent width in spectra which do not extend far into the red so that sampling this side of the continuum is difficult.  For this reason, we set larger upper error bars on these estimates and provide lower limits as the lower error bars by fitting the continuum immediately at the edge of the intermediate-width H$\alpha$ component on the red side.  We see a general trend of increasing equivalent width in the H$\alpha$ line until day $\sim 150$.

Though the error bars are rather large owing to variable slit widths and weather conditions, there is a general trend of increasing flux until about day 150 in the broad component, primarily because of increases in flux on the redshifted side of the line.  This is likely due to one of three situations.  Either the near side has become increasingly optically thin so that the far side is no longer occulted by material approaching us, the material on the far side has reached a distance great enough that it is now outside of the occultation region from our line of sight, or the SN ejecta on the far side are running into CSM for the first time.  

Eventually, the H$\alpha$ flux begins to decline, though evidence of weaker CSM interaction persists even in the day 1206 spectrum when the intermediate-width component of H$\alpha$ can still be seen.  As the host-galaxy light begins to dominate the total flux, the equivalent width shown in Figure \ref{fig:HalphaEWFluxhigh} drops.

Figure \ref{fig:HalphaTripleCompare} shows the H$\alpha$ line-profile evolution.  The most interesting evolution is seen as the broad component changes with time, though we also discuss the intermediate-width component later in this section.  Our earlier spectrum  on day 16 shows a broad profile with an extended blue wing in emission out to $-14{,}000$\,km\,$\mathrm{s^{-1}}$ at zero intensity (though the line strength is weak relative to the continuum at these early times).  The line appears truncated on the red side, perhaps due to occultation of the far side on days 16 and 30.  Because the broad blue component is first seen in the day 16 spectra and is measured at $-14{,}000$\,km\,$\mathrm{s^{-1}}$ at zero intensity, we can estimate the radius of the CSM on day 16 to be $R = vt \approx 130$\,au.  We also estimate the photospheric radius ($R = \sqrt{L/(4 \pi \sigma T^{4})} = 76^{+60}_{-27}$\,au).  Taking into account the large uncertainties in the temperature estimates from our blackbody fits ($T_{\rm eff} = 12{,}000$\,K $\pm\, 3{,}000$\,K) and the time until first CSM interaction, our radius estimates are consistent with the photosphere originating in the CSM interaction region at early times.

\begin{figure}
\centering
\includegraphics[width=0.5\textwidth,height=0.5\textheight,keepaspectratio,clip=true,trim=0cm 0cm 0cm 0cm]{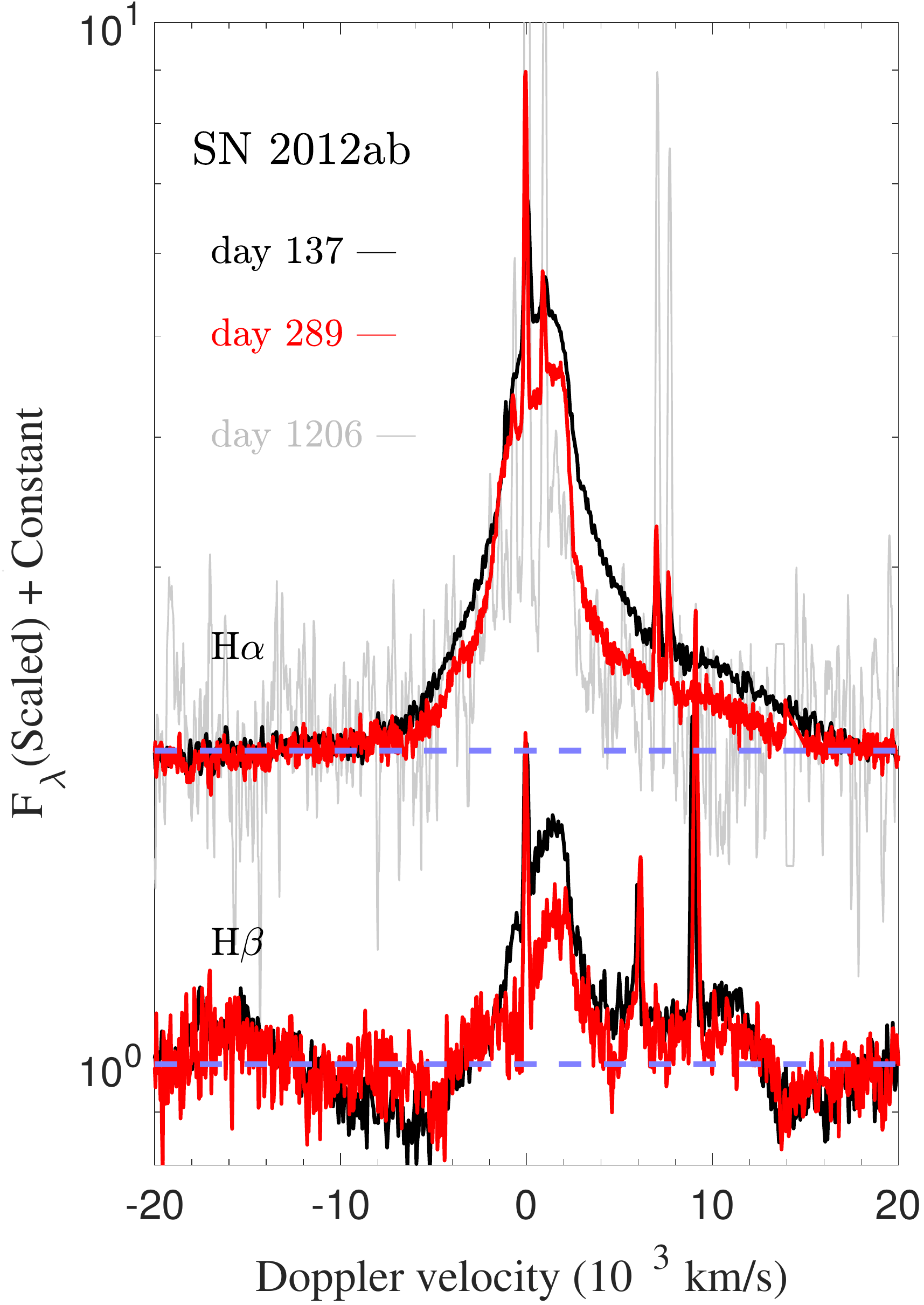}
\caption{H$\alpha$ and H$\beta$ in the late-time spectra of SN~2012ab.  See \S \ref{sec:Dis:asymHalpha} for a discussion of the broad redshifted component of H$\alpha$ and H$\beta$ that we see in the late-time spectra.}
\label{fig:Halphabetalate}
\end{figure}

Over time this asymmetric profile changes dramatically.  The broad blue wing at $-14{,}000$\,km\,$\mathrm{s^{-1}}$ diminishes and the blue edge slows to below $-7{,}000$\,km\,$\mathrm{s^{-1}}$ at zero intensity by day 119.  We see a more symmetric line core appear in the spectra on days 76 and 77.  By day 137, we clearly detect an augmented broad red component of the H$\alpha$ line out to $+20{,}000$\,km\,$\mathrm{s^{-1}}$ at zero intensity (see Figures \ref{fig:HalphaTripleCompare} and \ref{fig:Halphabetalate}).  By day 289 this broad red component of the H$\alpha$ line has slowed to about $+15{,}000$\,km\,$\mathrm{s^{-1}}$ at zero intensity.  The H$\beta$ emission profile also shows a similar broad red component, though we cannot determine its full extent because of blending with other spectral features on the red side.  

We show the late-time asymmetric broad profiles of H$\alpha$ and H$\beta$ more clearly in Figure \ref{fig:Halphabetalate}.  This broad red wing may be present in some of the earlier spectra as well, though it is difficult to detect because some of the spectra do not extend far enough to effectively constrain the continuum level on the red side (Figure \ref{fig:allspectra}).  While an augmented blue component can be caused by dust formation blocking the redshifted side of the line, an augmented red component requires real geometrical asphericity.  Since the broad red component at late times extends out to $+20{,}000$\,km\,$\mathrm{s^{-1}}$, this fast material has not yet come into contact with much, if any, CSM prior to day 137. The most likely origin for this broad red component is freely expanding SN ejecta  that are crossing the reverse shock \citep[see, e. g.,][]{2005ApJ...635L..41S}.  Consequently, we can estimate a minimum radius for the CSM interaction region (located outside of the reverse shock radius) of $R = vt \approx 1{,}600$\,au by assuming a constant SN ejecta velocity of $20{,}000$\,km\,$\mathrm{s^{-1}}$ until day 137.  

We also see distinct changes in the intermediate-width component of H$\alpha$ (Figure \ref{fig:HalphaTripleCompare}).  Initially, the intermediate-width component is blended with the broad emission.  By day 119, we detect a very prominent intermediate-width component in the H$\alpha$ profile, suggesting interaction with CSM.  At day 137, this intermediate-width component extends out to $-1{,}400$\,km\,$\mathrm{s^{-1}}$ at half maximum on the blue side and $+2{,}700$\,km\,$\mathrm{s^{-1}}$ on the red side.  It remains strong at a similar $-1{,}500$\,km\,$\mathrm{s^{-1}}$ at half maximum on the blue side and $+2{,}500$\,km\,$\mathrm{s^{-1}}$ at half maximum on the red side until day 289, and may even be present in our latest spectrum on day 1{,}206.  Much like the broad component of the H$\alpha$ emission at late times, we see an augmented red side to the intermediate-width profile, suggesting asphericity in the density of the emitting material.  The intermediate-width component at late times most likely arises from the post-shock swept-up CSM in the cold dense shell on the far side.  This is the same gas with which the reverse-shocked ejecta are interacting.

\begin{figure}
\centering
\includegraphics[width=0.5\textwidth,height=0.5\textheight,keepaspectratio,clip=true,trim=0cm 0cm 0cm 0cm]{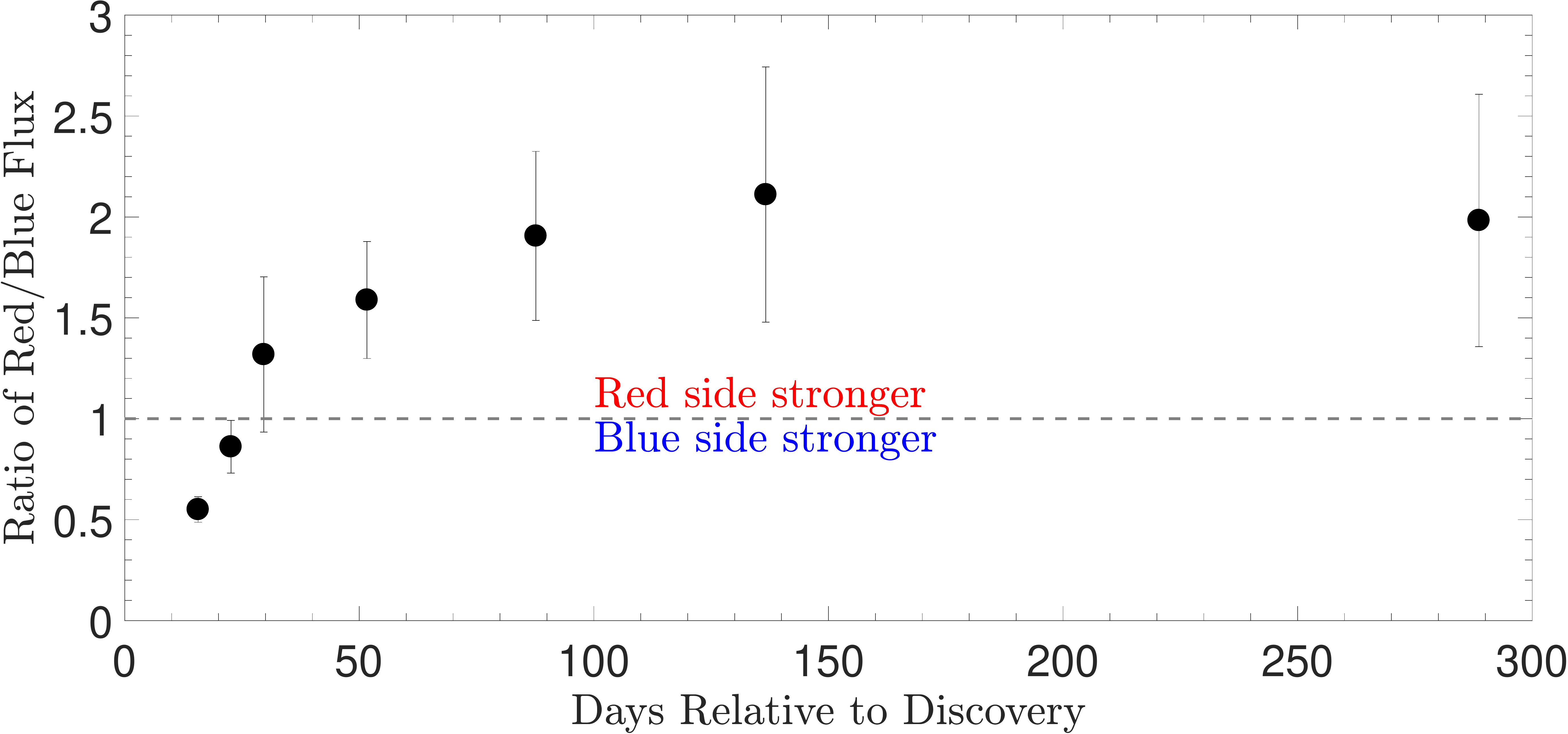}
\caption{Ratio of the equivalent width of the redshifted H$\alpha$ component (excluding the narrow component) to that of the blueshifted component.  We exclude all spectra which show no broad H$\alpha$ profile or have insufficient data on the redshifted side to reliably fit the continuum.}
\label{fig:ratiored}
\end{figure}

We measure the flux of the H$\alpha$ line above the continuum level (excluding the narrow component) on the redshifted and blueshifted sides of the line separately (choosing $v=0$ at the centre of the narrow component for each spectrum).  The ratio of these red/blue fluxes is shown in Figure \ref{fig:ratiored}.  The story here is consistent with that shown in Figure \ref{fig:HalphaTripleCompare}; we find an initially dominant blueshifted H$\alpha$ line that is quickly surpassed in strength by the redshifted H$\alpha$ component, which remains strong until late times.  We further discuss the implications of the changing H$\alpha$ line shape on the physical picture we construct for SN~2012ab in \S \ref{sec:Dis:picture}.

\subsection{Sources of Polarization}

As stated in \S \ref{sec:Res:Specpol}, we measure continuum polarization as high as $3.5\%\pm0.1\%$ in our MMT data.  Because the interstellar polarization level is relatively small ($<0.70\%$), this continuum polarization is most likely caused by large asphericity in the electron-scattering photosphere of SN~2012ab.  Since most of the flux of SN~2012ab on days 52--76 comes from CSM interaction, we expect the electron-scattering photosphere to arise primarily in the CSM interaction regions.  Although it is difficult to disentangle geometry, optical-depth effects, inclination angle, and a variety of other factors that affect the polarization signal we receive, SN~2012ab must have a heavily aspherical CSM density profile to produce such large continuum polarization values.  \citet{2015MNRAS.449.4304D} modeled SN~2010jl, another SN~IIn, with an axially symmetric prolate morphology and found that a pole-to-equator density ratio of $\sim 2.6$ could effectively produce the $\sim 1.6\%$ continuum polarization seen in spectropolarimetry of SN~2010jl.  As shown later (Figure \ref{fig:compare}), spectra of SN~2012ab are qualitatively similar to those of SN~2010jl.  

CSM dust outside of the interaction regions could affect our polarization signal.  We observe an increase in polarization between days 52 and 76 with no significant change in position angle.  This increase in polarization may be partly caused by a combination of an aspherical electron-scattering region and an external dusty CSM that is distributed in the same geometry.  This effect was seen in SN~2013ej \citep{2017ApJ...834..118M}.

\subsection{Depolarization of H$\alpha$ and H$\beta$}

Depolarization of H$\alpha$ and H$\beta$ is seen in our spectropolarimetry taken with the MMT on day 76.  The depolarization seen on day 76 is likely due to dilution by unpolarized line emission.  The earlier spectropolarimetry from the Bok telescope on day 52 does not show significant depolarization, though there may be some.   This suggests that the H$\alpha$ line may be influenced by electron scattering in addition to unpolarized line emission.  This is supported by the fact that the relative strength of the intermediate-width component of H$\alpha$ greatly increases between days 52 and 76, alongside the increase in the depolarization in H$\alpha$ seen between days 52 and 76.

Figure \ref{fig:QUcircle} displays the evolution of the day 52 and day 76 spectropolarimetry in the $q$--$u$ plane.  On both days we see the data form a locus of points along a line approaching the origin.  While we cannot disentangle the shape of the SN ejecta and the CSM from the position angle, the depolarization of the H$\alpha$ and the H$\beta$ lines along a linear region in the $q$--$u$ plane does suggest some level of axisymmetry in the emission regions.  We find that a disk-like geometry for the CSM, as has been suggested previously for some SNe~IIn, is plausible \citep{2006Natur.440..505L,2008ApJ...688.1186H,2014MNRAS.442.1166M,2015MNRAS.449.1876S}.

\subsection{Comparison to other SNe~II}
\label{sec:Dis:compareSNeII}
The light curve of SN~2012ab shows similarities to those of some other SNe~II (particularly the brightness and shape of SN~1998S), but the spectral evolution is unusual.  Our first spectrum was taken 7 days after discovery.  We cannot reliably measure the nebular lines in this spectrum because of blurring with nearby lines owing to its low resolution, which makes determination of the source of the narrow H$\alpha$ emission difficult on this date.  Additionally, the characteristic narrow lines associated with SNe~IIn can sometimes fade quickly (e.g., SN~1998S; \citealp{2015ApJ...806..213S}, PTF11iqb; \citealp{2015MNRAS.449.1876S}), so we may have missed this narrow CSM emission owing to a lack of spectra at very early times.  Since the very narrow component of H$\alpha$ emission ($< 300$\,km\,$\mathrm{s^{-1}}$) is likely due to H~II regions or distant CSM, we cannot use this component of the spectrum to classify SN~2012ab as a SN~IIn.  However, we do see clear evidence for CSM interaction in SN~2012ab in the intermediate-width H$\alpha$ emission.  

We also consider SN Types II-L and II-P for comparison.  There are two clues that CSM interaction in SN~2012ab is stronger than in normal SNe~II-P and SNe~II-L. (1) There is no P Cygni absorption profile seen in H$\alpha$, suggesting that while the pre-shock CSM around SN~2012ab may not be dense enough to produce narrow optical emission lines, it can produce X-ray emission which heats the interior SN ejecta so that we do not see absorption \citep{1994ApJ...420..268C}.  (2) The strong intermediate-width component arising by day 76 is likely the result of a swept-up CSM shell. 

Some SNe~IIn with relatively weak CSM interaction have been shown to resemble SNe~II-L (PTF11iqb; \citealp{2015MNRAS.449.1876S}).  SNe~II-L are a small fraction of the total SN population \citep{2011MNRAS.412.1522S}.  It has been proposed recently that there may be a continuum between the II-P and the II-L classification types \citep{2014ApJ...786...67A,2016MNRAS.459.3939V}, though other work suggests that they are two distinct populations \citep{2012ApJ...756L..30A}.  \citet{2009ApJ...694.1067P} argued that SNe~II-P should have a plateau phase that lasts roughly 100 days.  In a similar fashion, \citet{2014MNRAS.445..554F} suggested that a decline of less than 0.5\,mag by 50 days after peak is indicative of a SN~II-P.  Our photometry does not show a plateau phase lasting 100 days, but it also does not drop by 0.5\,mag in the first 50 days, making it less likely that SN~2012ab is a SN~II-P or a weakly interacting SN~IIn that resembles a SN~II-L.

\begin{figure*}
\centering
\includegraphics[width=1\textwidth,clip=true,trim=0cm 0cm 0cm 0cm]{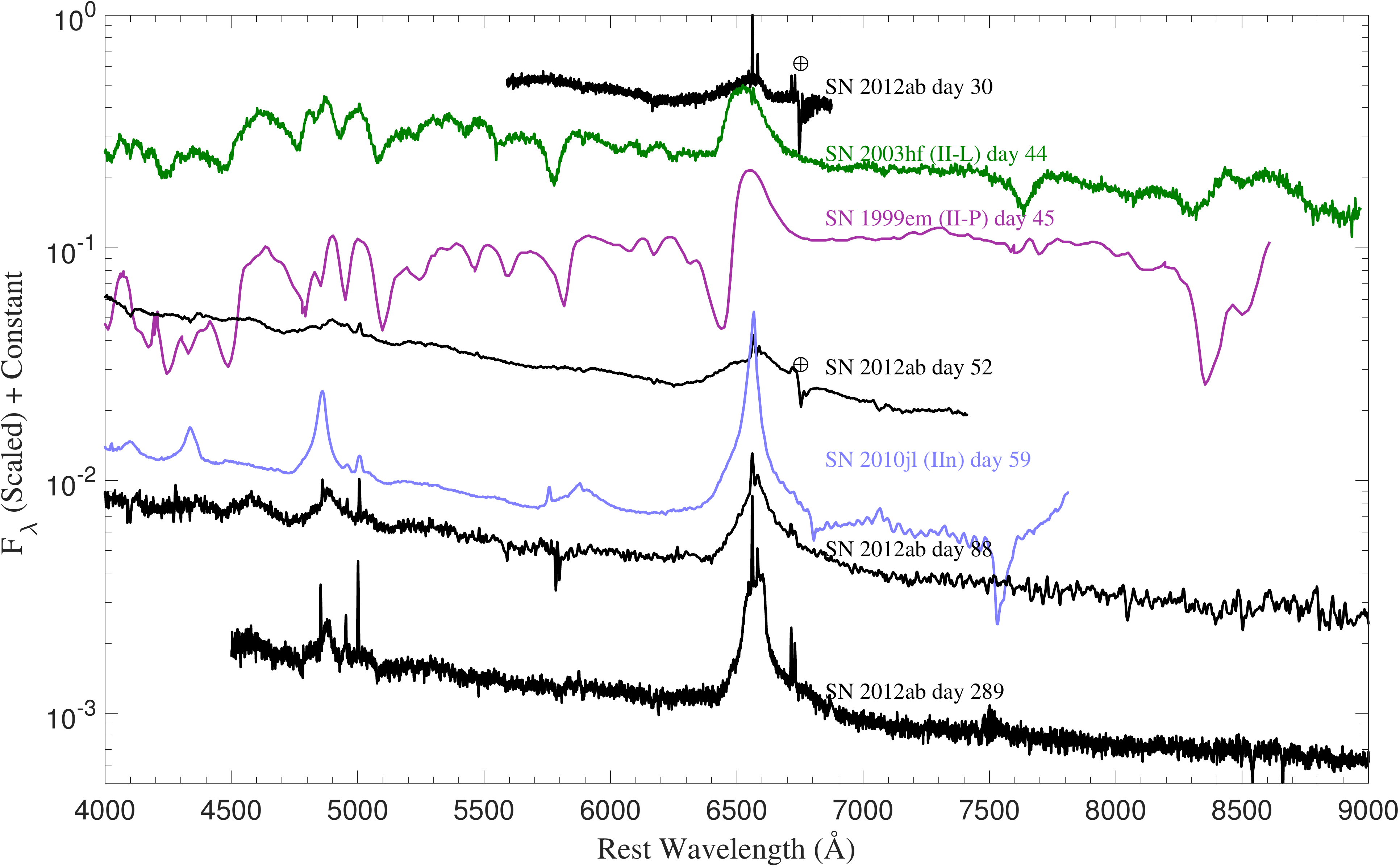}
\caption{Comparison of the SN~2012ab spectra to spectra of a variety of SNe [IIn: SN 2010jl \citep{inprep}; II-P: SN 1999em \citep{2002PASP..114...35L}; II-L: SN 2003hf \citep{2014MNRAS.445..554F}] at similar times.  SN~2012ab clearly lacks the strong absorption seen in the SN~II-P spectrum, the strong Ca component seen in the SN~IIn and SN~II-L spectra, and the various other metallic absorption lines blueward of H$\alpha$ seen in the SN~IIn and SN~II-L spectra.}
\label{fig:compare}
\end{figure*}

SNe~II-L generally do not show strong P Cygni absorption in the H$\alpha$ line profile compared to SNe~II-P \citep{1996AJ....111.1660S}.  However, SNe~II-L still tend to have a number of Fe \textsc{ii}, Ba \textsc{ii}, Sc \textsc{ii}, Mg \textsc{i}, Ti \textsc{ii}, and Ca \textsc{ii} absorption or P Cygni features in the visible wavelength range, which SN~2012ab lacks.  An underlying SN~II-P can look like a SN~II-L because of extra luminosity from CSM interaction and CSM interaction modifying the spectrum, as appeared to be the case for PTF11iqb and a few other objects \citep{2015MNRAS.449.1876S,2016arXiv161008054M}.  In SN~2012ab, these features in the underlying SN atmosphere may be masked to an even greater extent because of stronger CSM interaction.  Figure \ref{fig:compare} shows a comparison of select epochs of the SN~2012ab spectra to those of well-known SNe~IIn, II-L, and II-P.  The SN~2012ab spectra are similar to that of SN~2010jl on day 59.  Although the intermediate-width H$\alpha$ component seen in SN~2012ab is indeed somewhat broader than is typically seen in other Type IIn SNe, these intermediate-width lines are uncharacteristic of other Type II SNe.  While SN~2012ab shows some similarities to SNe~II-L but with stronger CSM interaction, it is most clearly a Type IIn SN because of its intermediate-width lines and relatively smooth continuum.

\subsection{A Physical Picture for SN~2012ab}
\label{sec:Dis:picture}
\begin{figure*}
\centering
\includegraphics[width=0.8\textwidth,clip=true,trim=0cm 0cm 0cm 0cm]{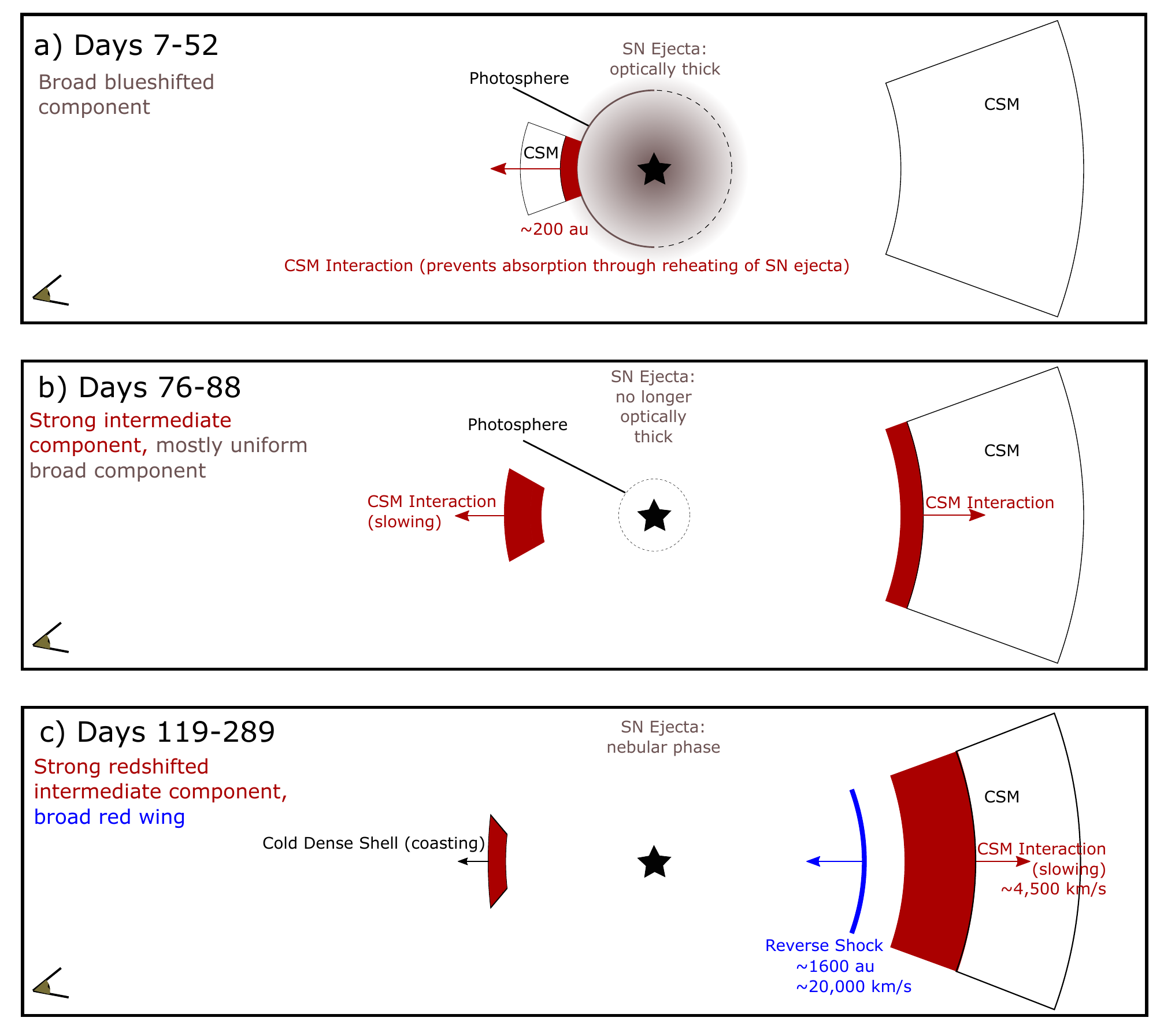}
\caption{Overview of the evolution of SN~2012ab.  Distances in the panels are estimated based on the ejecta velocities described in the text.  The viewing angle on the left side is not constrained well, but is probably out of the equatorial plane owing to a lack of narrow absorption seen in the spectra.  We see a general change from prominent blueshifted H$\alpha$ emission at early times to prominent redshifted H$\alpha$ emission at late times, likely caused by asphericity in the CSM.  A more detailed description of each panel is given in the text in \S \ref{sec:Dis:picture}.}
\label{fig:cartoon}
\end{figure*}

Figure \ref{fig:cartoon} provides a schematic overview and description of the possible physical evolution of SN~2012ab.  There are three main stages of evolution captured by our spectra, as follows.

\begin{enumerate}
	\item The primary feature in the early-time spectra of SN~2012ab (days 7--52, Figure \ref{fig:cartoon}a) is a broad blue wing of H$\alpha$ extending to $-14{,}000$\,km\,$\mathrm{s^{-1}}$ with a truncated red wing.  This could be caused by a very large asphericity in the explosion of SN~2012ab, an optically thick photosphere blocking emission from redshifted material, or a lack of CSM interaction on the red side.  Since we see redshifted emission in both an intermediate-width and a broad component at later times, there must be fast ejecta and CSM on the receding side.  Very broad redshifted emission at late times indicates that some of the SN ejecta have traveled until those late times without yet crashing into any CSM.  Additionally, because the line wings are asymmetric with a broader blue wing at first and a broader red wing later, they cannot be due exclusively to broadening by electron scattering.

Taking all of this into account, we favour the interpretation that initially the photosphere is optically thick and blocks emission from most of the redshifted ejecta, while there is little or no CSM interaction on the far side.  The narrow emission lines seen in H$\alpha$ at early times may be H~II region lines or distant CSM.  We do not see any narrow absorption features, so we suspect that our viewing angle is out of the equatorial plane of the CSM interaction.  Alternatively, the lack of narrow P Cygni features in the spectra suggests that the near-side CSM interaction depicted in Figure \ref{fig:cartoon} may be enveloped by the SN photosphere along our line of sight, as was suggested for SN~IIn PTF11iqb \citep{2015MNRAS.449.1876S}.  A lack of higher-velocity absorption features from the fast-moving SN ejecta suggests that the shocked CSM is likely emitting X-rays and thus reheating the SN ejecta.
	
	\item The primary features in the intermediate-phase spectra of SN~2012ab (days 76--88, Figure \ref{fig:cartoon}b) are a strong intermediate-width component and a mostly symmetric broad component.  Because the initially obscuring blueshifted material has expanded and cooled, causing the photosphere to recede, we can now see a redshifted H$\alpha$ component in the spectra of SN~2012ab.  This redshifted H$\alpha$ consists of both an intermediate-width and a broad component.  The broad blue wing of H$\alpha$ has gradually slowed down as the fastest ejecta have already run into more CSM, creating an intermediate-width component to the blueshifted H$\alpha$ line.  Initially, the intermediate-width line core is roughly symmetric, but with time we see that the red side of the line has both a greater velocity and brightness.	

	\item The primary features in the late-time spectra of SN~2012ab (days 119--289, Figure \ref{fig:cartoon}c) are a strong redshifted intermediate-width component and a very broad red wing with relatively weak emission on the blue side (see Figure \ref{fig:Halphabetalate}).  The elevated brightness of the red side suggests a higher density of CSM on the receding side of SN~2012ab.  On the other hand, a higher density of material on the red side would generally mean more slowing of the SNe ejecta, but we see a higher velocity on the red side.  This implies that the CSM on the red side had an initially larger distance away from the SN, so interaction with the SN ejecta began at a later time and the fast ejecta had not yet decelerated.

The most unusual aspect of SN~2012ab is its broad red wing of H$\alpha$ at $+20{,}000$\,km\,$\mathrm{s^{-1}}$, which is clearly present on day 137.  \citet{2009MNRAS.394...21D} show that electron scattering in high velocity ejecta can produce an asymmetric broad red wing, while the associated P Cygni absorption on the blue side is not always seen in Type IIn SNe.  Although electron scattering may contribute to the extended red wing in SN~2012ab, it cannot explain the augmented red intermediate-width component nor is it clear if it can produce the broadest part of the red wing out to $+20{,}000$\,km\,$\mathrm{s^{-1}}$.  Because there is not a similar blue wing at the same time, we conclude that the broadest part of the red wing is likely caused by asphericity in the density and radius of the CSM.  It is likely that this bright and broad H$\alpha$ emission at such late times arises from the reverse shock, as has also been seen in SN~1987A \citep{2005ApJ...635L..41S,2011ApJ...743..186F,2013ApJ...768...88F}.  In the case of SN~2012ab, this material could not previously have crashed into any closer CSM and is estimated to be at a distance of $\sim 1600$\,au from the central SN (see \S \ref{sec:Dis:asymHalpha}).  The fact that we also observe a stronger red side to the intermediate-width component of the H$\alpha$ line at late times reinforces the idea that on the far side of SN~2012ab the CSM interaction begins later, the CSM has higher density at that radius, and the SN ejecta have higher velocities because they have not decelerated yet.  Since the late-time broad red wing originates in material traveling $+20{,}000$\,km\,$\mathrm{s^{-1}}$, the redshifted SN ejecta were expanding unimpeded in a relative cavity at smaller radii.
\end{enumerate}

\subsection{Implications}
\label{sec:Dis:implications}
The high luminosity of SN~2012ab, its intermediate-width spectral features, and the lack of normal P Cygni features in the ejecta photosphere suggest that it is undergoing CSM interaction.  Both the spectral evolution and the spectropolarimetry provide evidence that this CSM is highly aspherical.  While we have little direct information about the nature of the progenitor of SN~2012ab (our limiting magnitudes only go to a depth of $M = -16.1$\,mag at best and span 3 days prior to discovery), the aspherical CSM provides valuable information about the mass-loss history of SN~2012ab.  In order to produce lower-density CSM closer to SN~2012ab on the side facing us, and higher-density CSM at larger radius on the far side, SN~2012ab must have undergone variable and aspherical mass loss.  A spherical wind from a single star and axisymmetric CSM from a rotating star both fail to explain the asymmetric spectral features that we see.  Instead, we speculate that such an aspherical CSM may have been the result of interactions analogous to the colliding winds seen in pinwheel nebulae \citep{1999Natur.398..487T,1999ApJ...525L..97M,2004MNRAS.349.1093R,2009A&A...506L..49M} or interactions in a highly eccentric binary system like $\eta$ Carinae, producing different amounts of CSM in each direction in the equatorial plane.  These possibilities are explored below.

A particularly interesting scenario resulting in nonaxisymmetric mass loss is that of a WR star and an O (or B) star in a binary system where their winds collide and produce a ``pinwheel'' nebula \citep{1999Natur.398..487T}.  These have mass-loss rates of $10^{-6}\, {\rm M}_{\odot}\,\mathrm{yr^{-1}}$ in hydrogen-poor WR winds \citep{2009A&A...506L..49M}.  In SN~2012ab, however, we require much higher mass-loss rates of hydrogen-rich material, but it is interesting to think if a similar binary interaction may be operating.  If oriented properly, the resulting spiral-like structure might allow the SN ejecta to crash into CSM on the near and far sides at different times, producing the lopsided broad signature that we see in our spectra.  We can estimate the period of the hypothetical binary orbit based on the times when we see CSM interaction begin on the close compared to the far sides.  We see CSM interaction early in our spectra on day 7 on the blueshifted side, and the redshifted CSM interaction does not turn on until about day 76.  

Since the high-velocity SN ejecta traveling at $+20{,}000$\,km\,$\mathrm{s^{-1}}$ took $\sim 70$ days longer to reach the CSM on the far side than the close side, we can extrapolate that the original wind, which was traveling at a lower velocity of several hundred km\,$\mathrm{s^{-1}}$ at most, must have been ejected on the far side several thousands of days prior to the material that was being ejected toward us.  This gives us an important constraint on the period of our likely binary progenitor system.  The period we estimate is longer than those of WR~118 ($P \approx 60$\,days; \citealp{2009A&A...506L..49M}), WR~104 ($P=220$\,days; \citealp{1999Natur.398..487T}), WR~98 ($P=565$\,days; \citealp{1999ApJ...525L..97M}), and the expected WR star progenitor to the Type IIb SN~2001ig ($P\approx 150$\,days; \citealp{2004MNRAS.349.1093R}).  Although mass loss from pinwheel nebulae might produce the aspherical spectral features like the ones we see, the periods of these systems are much shorter than the time delay we see in our data, suggesting that a circularised binary system is not a likely candidate for the progenitor of SN~2012ab.

Other candidates for such mass-loss events include WR stars and LBVs that are undergoing heavy mass loss and are in highly eccentric orbits.  Pairing these kinds of objects into an elliptical binary orbit, especially with other massive stars, can result in colliding stellar winds and the eventual production of dust \citep{2014MNRAS.445.1663Z,2014ApJ...785....8Z,2015PASJ...67..121S}.  One such system, WR~140, has a high eccentricity of $e = 0.881$ and period of $P = 2{,}899$\,days \citep{2003ApJ...596.1295M}, resulting in a very eventful periastron passage with a resolved one-sided nebula \citep{2002ApJ...567L.137M}.  Another system with a qualitatively similar colliding wind is $\eta$ Carinae, a well-studied LBV binary system \citep{1996ApJ...460L..49D} with eccentricity $e =0.9$ and period $P = 5.5$\,yr \citep{2001ApJ...547.1034C,2009MNRAS.394.1758P}.  At the present epoch, $\eta$ Carinae forms dust episodically in its colliding stellar wind \citep{2010MNRAS.402..145S}, much like WR~140 but at a higher mass-loss rate and lower wind velocity of $\sim 500$\,km\,$\mathrm{s^{-1}}$ \citep{2001ApJ...553..837H}.  $\eta$ Carinae is also a more appropriate comparison for SN~2012ab because it has a hydrogen-rich wind, unlike the wind from a hydrogen-poor WR star.

\citet{2016MNRAS.463..845K} show that $\eta$ Carinae has exhibited strong nonaxisymmetric ejection in its three periods of eruptive mass loss over the past 1{,}000\,yr.  Although $\eta$ Carinae's 1840s mass-loss eruption was mostly axisymmetric, its two prior eruptions in the 13th and 16th centuries were not, suggesting that the mass-loss mechanism is highly variable and can even be one-sided\footnote{Recent ALMA observations show a highly nonaxisymmetric density distribution in the equatorial material around $\eta$ Carinae, even in the otherwise axisymmetric 19th century eruption \citep{SmithALMAsubmitted}}.  Our estimated lower bound on the mass-loss rate for SN~2012ab of $\dot{M} = 0.050\,{\rm M}_{\odot} \mathrm{yr^{-1}}$ is very high for what would be expected from a WR star, although plausible for an LBV-like $\eta$ Carinae during an eruption.  Moreover, the H-rich CSM in SN~2012ab is clearly inconsistent with a normal WR star, pointing instead toward an LBV.  

Binary systems have been suggested to play a critical role in other work on SN~IIn progenitors \citep{2013MNRAS.436.2484K,2014ApJ...785...82S,2014ARA&A..52..487S,2014MNRAS.438.1191S,2014MNRAS.442.1166M} and LBV systems undergoing heavy mass loss \citep{2011MNRAS.415.2020S,2015MNRAS.447..598S}.  Because of the sheer magnitude of the mass-loss rates and aspherical nature we see in SN~2012ab, it is more likely that its progenitor system is a binary with an LBV undergoing eruptive mass loss than any of the WR binary scenarios we discussed.  75\% of high-mass stars are likely to undergo binary interaction during their lifetimes \citep{2012Sci...337..444S}, and binary stars with longer periods tend to have higher eccentricities \citep{1991A&A...248..485D}, resulting in stronger interactions at periastron passage .  It is plausible, then, that SN~2012ab may be the result of a high-mass wide binary star system undergoing interaction during periastron passage at some point prior to exploding as a SN.

\section{Summary}
SN~2012ab bears many similarities to other SNe with CSM interaction.  Its spectrum is very similar to that of SN~2010jl on day 59 and its light curve is very similar to that of SN~1998S.  SN~2012ab does, however, exhibit several properties that distinguish it from other SNe~IIn, as follows.

\begin{itemize}

 \item A quick rise time of $\lesssim 3$\,days suggests that we are seeing the sudden onset of CSM interaction.  The SN ejecta may have been expanding undetected in an inner cavity prior to this interaction or it may have taken a few days before we could see the shock breakout in the wind.
 \item The broad blue wing of H$\alpha$ extending to $-14{,}000$\,km\,$\mathrm{s^{-1}}$ at early times suggests there are fast-moving ejecta on the near side of the SN, and that the SN ejecta are initially optically thick.
 \item The lack of the broad red wing of H$\alpha$ at early times and its presence at $+20{,}000$\,km\,$\mathrm{s^{-1}}$ at late times suggests an initial occultation of the far side, combined with a later start time for CSM interaction on the receding side of SN~2012ab.  The delay in CSM interaction on the receding side is most likely caused by differences in the radius of the CSM.  This requires aspherical mass loss with the material ejected toward us being released at a more recent time (or lower velocity) and the material ejected away from us being released farther back in time (or at a higher velocity).
 \item The intermediate-width line core of H$\alpha$ has a larger brightness and higher velocity on the receding side of SN~2012ab at later epochs, suggesting a higher density of heated post-shock material at late times.  The CSM  producing the blue side of the intermediate-width line is likely already swept up and has cooled at this point, so we see a resulting discrepancy in brightness compared to the red side.
 \item The relatively high continuum polarization level at 5{,}400--5{,}500\,{\AA} of $1.7\%\pm0.1\%$ increases to an even higher value of $3.5\%\pm0.1\%$ by day 76.  Strong depolarization of the H$\alpha$ line seen between days 52 and 76 occurs at the same time that we see a prominent intermediate-width component emerge in H$\alpha$, indicating that the continuum polarization is not caused primarily by scattering off of CSM dust.  This unpolarized line emission from CSM interaction dominates the light in the H$\alpha$ line, thus providing additional support to the idea that the continuum polarization comes from real asphericity in the SN event.
 \item We derive a likely progenitor mass-loss rate of $\dot{M} = 0.050\,{\rm M}_{\odot}\,\mathrm{yr^{-1}}$, assuming a wind speed of $v_{\rm pre} = 100$\,km\,$\mathrm{s^{-1}}$.  This heavy mass-loss rate needs to be highly aspherical to explain the asymmetric line profiles, which we suggest may have arisen from nonaxisymmetric mass loss in an eccentric binary system.  The asymmetric line profiles do not require mass loss in the form of a steady wind.  Instead, a brief ejection which creates a dense shell would suffice.
\end{itemize}

\section*{Acknowledgments}
This work was supported by NSF grant AST-1210599.  N.S. received additional support from NSF grants AST-1312221 and AST-1515559, and by a Scialog grant from the Research Corporation for Science Advancement. J.M.S. was supported by an NSF Astronomy and Astrophysics Postdoctoral Fellowship under award AST-1302771.  The work of A.V.F.'s supernova group at UC Berkeley has been generously supported by NSF grant AST-1211916, the TABASGO Foundation, Gary and Cynthia Bengier, the Christopher R. Redlich Fund, the Richard \& Rhoda Goldman Fund, and the Miller Institute for Basic Research in Science (U.C. Berkeley).  J.C.W.'s SN group at UT Austin was supported by NSF grant AST-1109881.  ROTSE-III was supported by NASA grant NNX-08AV63G, NSF grant PhY-0801007, the Australian Research Council, the University of New South Wales, the University of Texas, and the University of Michigan.  D.C.L. acknowledges support from NSF grants AST-1009571 and AST-1210311, under which part of this research was carried out.

This research used astrometric solutions from Astrometry.net and data provided by the USNO-B Image and Catalogue Archive operated by the United States Naval Observatory. We also made use of the NASA/IPAC Extragalactic Database (NED), which is operated by the Jet Propulsion Laboratory, California Institute of Technology, under contract with NASA.

We thank the staffs at the MMT, Bok, Lick, and Keck Observatories for their assistance with the observations.  Observations using Steward Observatory facilities were obtained as part of the large observing program AZTEC: Arizona Transient Exploration and Characterization. Some observations reported here were obtained at the MMT Observatory, a joint facility of the University of Arizona and the Smithsonian Institution.  Some of the data presented herein were obtained at the W. M. Keck Observatory, which is operated as a scientific partnership among the California Institute of Technology, the University of California, and NASA; the observatory was made possible by the generous financial support of the W.M. Keck Foundation. The authors wish to recognise and acknowledge the very significant cultural role and reverence that the summit of Maunakea has always had within the indigenous Hawaiian community. We are most fortunate to have the opportunity to conduct observations from this mountain.  Research at Lick Observatory is partially supported by a generous gift from Google.

The Hobby-Eberly Telescope (HET) is a joint project of the University of Texas at Austin, the Pennsylvania State University, Stanford University, Ludwig-Maximilians-Universit{\"a}t M{\"u}nchen, and Georg-August-Universit{\"a}t G\"{o}ttingen. The HET is named in honor of its principal benefactors, William P. Hobby and Robert E. Eberly.  We acknowledge the HET Resident Astronomer Team (M. Shetrone, S. Odewahn,
J. Caldwell, S. Rostopchin) for their work and support during the observations taken at McDonald Observatory.

\appendix

\bsp

\label{lastpage}


\begin{thebibliography}{99}
\bibitem[Abazajian et al.(2009)]{2009ApJS..182..543A} Abazajian, K.~N., Adelman-McCarthy, J.~K., Ag{\"u}eros, M.~A., et al.\ 2009, \apjs, 182, 543 

\bibitem[Anderson et al.(2014)]{2014ApJ...786...67A} Anderson, J.~P., Gonz{\'a}lez-Gait{\'a}n, S., Hamuy, M., et al.\ 2014, \apj, 786, 67 

\bibitem[Arcavi et al.(2012)]{2012ApJ...756L..30A} Arcavi, I., Gal-Yam, A., Cenko, S.~B., et al.\ 2012, \apjl, 756, L30 

\bibitem[Arcavi et al.(2014)]{2014ApJ...793...38A} Arcavi, I., Gal-Yam, A., Sullivan, M., et al.\ 2014, \apj, 793, 38 

\bibitem[Arnett et al.(1989)]{1989ARA&A..27..629A} Arnett, W.~D., Bahcall, J.~N., Kirshner, R.~P., \& Woosley, S.~E.\ 1989, \araa, 27, 629 

\bibitem[Barron et al.(2008)]{2008AJ....135..414B} Barron, J.~T., Stumm, C., Hogg, D.~W., Lang, D., \& Roweis, S.\ 2008, \aj, 135, 414-422 

\bibitem[Bilinski et al.(2015)]{2015MNRAS.450..246B} Bilinski, C., Smith, N., Li, W., et al.\ 2015, \mnras, 450, 246  

\bibitem[Burrows et al.(1995)]{1995ApJ...450..830B} Burrows, A., Hayes, J., \& Fryxell, B.~A.\ 1995, \apj, 450, 830   

\bibitem[Chevalier \& Fransson(1994)]{1994ApJ...420..268C} Chevalier, R.~A., \& Fransson, C.\ 1994, \apj, 420, 268  

\bibitem[Chornock et al.(2010)]{2010ApJ...713.1363C} Chornock, R., Filippenko, A.~V., Li, W., \& Silverman, J.~M.\ 2010, \apj, 713, 1363 

\bibitem[Chugai(2001)]{2001MNRAS.326.1448C} Chugai, N.~N.\ 2001, \mnras, 326, 1448 

\bibitem[Chugai(2006)]{2006AstL...32..739C} Chugai, N.~N.\ 2006, Astronomy Letters, 32, 739 

\bibitem[Chugai \& Danziger(1994)]{1994MNRAS.268..173C} Chugai, N.~N., \& Danziger, I.~J.\ 1994, \mnras, 268, 173   

\bibitem[Chugai et al.(2005)]{2005AstL...31..792C} Chugai, N.~N., Fabrika, S.~N., Sholukhova, O.~N., et al.\ 2005, Astronomy Letters, 31, 792 

\bibitem[Corcoran et al.(2001)]{2001ApJ...547.1034C} Corcoran, M.~F., Ishibashi, K., Swank, J.~H., \& Petre, R.\ 2001, \apj, 547, 1034 

\bibitem[Damineli(1996)]{1996ApJ...460L..49D} Damineli, A.\ 1996, \apjl, 460, L49 

\bibitem[Dessart et al.(2009)]{2009MNRAS.394...21D} Dessart, L., Hillier, D.~J., Gezari, S., Basa, S., \& Matheson, T.\ 2009, \mnras, 394, 21 

\bibitem[Dessart \& Hillier(2011)]{2011MNRAS.410.1739D} Dessart, L., \& Hillier, D.~J.\ 2011, \mnras, 410, 1739 

\bibitem[Dessart et al.(2015)]{2015MNRAS.449.4304D} Dessart, L., Audit, E., \& Hillier, D.~J.\ 2015, \mnras, 449, 4304   

\bibitem[Duquennoy \& Mayor(1991)]{1991A&A...248..485D} Duquennoy, A., \& Mayor, M.\ 1991, \aap, 248, 485 

\bibitem[Evans \& Kochanek(1989)]{1989ApJ...346L..13E} Evans, C.~R., \& Kochanek, C.~S.\ 1989, \apjl, 346, L13 

\bibitem[Faber et al.(2003)]{2003SPIE.4841.1657F} Faber, S.~M., Phillips, A.~C., Kibrick, R.~I., et al.\ 2003, \procspie, 4841, 1657   

\bibitem[Faran et al.(2014a)]{2014MNRAS.445..554F} Faran, T., Poznanski, D., Filippenko, A.~V., et al.\ 2014a, \mnras, 445, 554   

\bibitem[Faran et al.(2014b)]{2014MNRAS.442..844F} Faran, T., Poznanski, D., Filippenko, A.~V., et al.\ 2014b, \mnras, 442, 844   

\bibitem[Fassia et al.(2000)]{2000MNRAS.318.1093F} Fassia, A., Meikle, W.~P.~S., Vacca, W.~D., et al.\ 2000, \mnras, 318, 1093   

\bibitem[Filippenko(1982)]{1982PASP...94..715F} Filippenko, A.~V.\ 1982, \pasp, 94, 715 

\bibitem[Filippenko(1997)]{1997ARA&A} Filippenko, A.~V.\ 1997, \araa, 35, 309 

\bibitem[France et al.(2011)]{2011ApJ...743..186F} France, K., McCray, R., Penton, S.~V., et al.\ 2011, \apj, 743, 186 

\bibitem[Fransson et al.(2005)]{2005ApJ...622..991F} Fransson, C., Challis, P.~M., Chevalier, R.~A., et al.\ 2005, \apj, 622, 991   

\bibitem[Fransson et al.(2002)]{2002ApJ...572..350F} Fransson, C., Chevalier, R.~A., Filippenko, A.~V., et al.\ 2002, \apj, 572, 350   

\bibitem[Fransson et al.(2014)]{2014ApJ...797..118F} Fransson, C., Ergon, M., Challis, P.~J., et al.\ 2014, \apj, 797, 118   

\bibitem[Fransson et al.(2013)]{2013ApJ...768...88F} Fransson, C., Larsson, J., Spyromilio, J., et al.\ 2013, \apj, 768, 88 

\bibitem[Gal-Yam(2012)]{2012Sci...337..927G} Gal-Yam, A.\ 2012, Science, 337, 927 

\bibitem[Gezari et al.(2012)]{2012Natur.485..217G} Gezari, S., Chornock, R., Rest, A., et al.\ 2012, \nat, 485, 217 

\bibitem[Hill et al.(1998)]{LRS-HET}Hill, G.J., Nicklas, H.E., MacQueen, P.J., Tejada, C., Cobos Duenas, F.J., and Mitsch, W. 1998, Proc. SPIE, 3355, 375

\bibitem[Hillier et al.(2001)]{2001ApJ...553..837H} Hillier, D.~J., Davidson, K., Ishibashi, K., \& Gull, T.\ 2001, \apj, 553, 837 

\bibitem[Hoffman et al.(2008)]{2008ApJ...688.1186H} Hoffman, J.~L., Leonard, D.~C., Chornock, R., et al.\ 2008, \apj, 688, 1186-1209 

\bibitem[H\"{o}flich(1991)]{1991A&A...246..481H} H\"{o}flich, P.\ 1991, \aap, 246, 481 

\bibitem[Kashi et al.(2013)]{2013MNRAS.436.2484K} Kashi, A., Soker, N., \& Moskovitz, N.\ 2013, \mnras, 436, 2484 

\bibitem[Katsuda et al.(2014)]{2014ApJ...780..184K} Katsuda, S., Maeda, K., Nozawa, T., Pooley, D., \& Immler, S.\ 2014, \apj, 780, 184   

\bibitem[Khokhlov et al.(1999)]{1999ApJ...524L.107K} Khokhlov, A.~M., H{\"o}flich, P.~A., Oran, E.~S., et al.\ 1999, \apjl, 524, L107 

\bibitem[Kiminki et al.(2016)]{2016MNRAS.463..845K} Kiminki, M.~M., Reiter, M., \& Smith, N.\ 2016, \mnras, 463, 845 

\bibitem[Leonard \& Filippenko(2001)]{2001PASP..113..920L} Leonard, D.~C., \& Filippenko, A.~V.\ 2001, \pasp, 113, 920 

\bibitem[Leonard \& Filippenko(2005)]{2005ASPC..342..330L} Leonard, D.~C., \& Filippenko, A.~V.\ 2005, 1604-2004: Supernovae as Cosmological Lighthouses, 342, 330   

\bibitem[Leonard et al.(2001)]{2001ApJ...553..861L} Leonard, D.~C., Filippenko, A.~V., Ardila, D.~R., \& Brotherton, M.~S.\ 2001, \apj, 553, 861 

\bibitem[Leonard et al.(2000)]{2000ApJ...536..239L} Leonard, D.~C., Filippenko, A.~V., Barth, A.~J., \& Matheson, T.\ 2000, \apj, 536, 239   

\bibitem[Leonard et al.(2002)]{2002PASP..114...35L} Leonard, D.~C., Filippenko, A.~V., Gates, E.~L., et al.\ 2002, \pasp, 114, 35   

\bibitem[Leonard et al.(2002)]{2002AJ....124.2490L} Leonard, D.~C., Filippenko, A.~V., Li, W., et al.\ 2002, \aj, 124, 2490 

\bibitem[Leonard et al.(2006)]{2006Natur.440..505L} Leonard, D.~C., Filippenko, A.~V., Ganeshalingam, M., et al.\ 2006, \nat, 440, 505   

\bibitem[Leonard et al.(2012)]{2012AIPC.1429..204L} Leonard, D.~C., Dessart, L., Hillier, D.~J., \& Pignata, G.\ 2012, American Institute of Physics Conference Series, 1429, 204 

\bibitem[Leonard et al.(2016)]{2016IAUFM..29B.458L} Leonard, D.~C., Dessart, L., Pignata, G., et al.\ 2016, IAU Focus Meeting, 29, 458 

\bibitem[Lodato \& Rossi(2011)]{2011MNRAS.410..359L} Lodato, G., \& Rossi, E.~M.\ 2011, \mnras, 410, 359 

\bibitem[Marchenko et al.(2003)]{2003ApJ...596.1295M} Marchenko, S.~V., Moffat, A.~F.~J., Ballereau, D., et al.\ 2003, \apj, 596, 1295 

\bibitem[Mauerhan et al.(2013)]{2013MNRAS.430.1801M} Mauerhan, J.~C., Smith, N., Filippenko, A.~V., et al.\ 2013, \mnras, 430, 1801 

\bibitem[Mauerhan et al.(2014)]{2014MNRAS.442.1166M} Mauerhan, J., Williams, G.~G., Smith, N., et al.\ 2014, \mnras, 442, 1166  

\bibitem[Mauerhan et al.(2017)]{2017ApJ...834..118M} Mauerhan, J.~C., Van Dyk, S.~D., Johansson, J., et al.\ 2017, \apj, 834, 118 

\bibitem[Maund et al.(2009)]{2009ApJ...705.1139M} Maund, J.~R., Wheeler, J.~C., Baade, D., et al.\ 2009, \apj, 705, 1139 

\bibitem[Miller \& Stone (1993)]{1993MillerStone} Miller, J. S., \& Stone, R. P. S. 1993, Lick Obs. Tech. Rep. 66 (Santa Cruz: Lick  Obs.)  

\bibitem[Miller et al. (1988)]{1988MillerSPOL} Miller, J. S., Robinson, L. B., \& Goodrich, R. W. 1988, Instrumentation for Ground-Based Optical Astronomy, 157  

\bibitem[Millour et al.(2009)]{2009A&A...506L..49M} Millour, F., Driebe, T., Chesneau, O., et al.\ 2009, \aap, 506, L49 

\bibitem[Monnier et al.(1999)]{1999ApJ...525L..97M} Monnier, J.~D., Tuthill, P.~G., \& Danchi, W.~C.\ 1999, \apjl, 525, L97 

\bibitem[Monnier et al.(2002)]{2002ApJ...567L.137M} Monnier, J.~D., Tuthill, P.~G., \& Danchi, W.~C.\ 2002, \apjl, 567, L137 

\bibitem[Morozova et al.(2016)]{2016arXiv161008054M} Morozova, V., Piro, A.~L., \& Valenti, S.\ 2016, arXiv:1610.08054 

\bibitem[Munari \& Zwitter(1997)]{1997A&A...318..269M} Munari, U., \& Zwitter, T.\ 1997, \aap, 318, 269   

\bibitem[Nicholl et al.(2015)]{2015MNRAS.452.3869N} Nicholl, M., Smartt, S.~J., Jerkstrand, A., et al.\ 2015, \mnras, 452, 3869 

\bibitem[Oke et al.(1995)]{1995PASP..107..375O} Oke, J.~B., Cohen, J.~G., Carr, M., et al.\ 1995, \pasp, 107, 375   

\bibitem[Parkin et al.(2009)]{2009MNRAS.394.1758P} Parkin, E.~R., Pittard, J.~M., Corcoran, M.~F., Hamaguchi, K., \& Stevens, I.~R.\ 2009, \mnras, 394, 1758 

\bibitem[Pastorello et al.(2013)]{2013ApJ...767....1P} Pastorello, A., Cappellaro, E., Inserra, C., et al.\ 2013, \apj, 767, 1 

\bibitem[Patat et al.(2011)]{2011A&A...527L...6P} Patat, F., Taubenberger, S., Benetti, S., Pastorello, A., \& Harutyunyan, A.\ 2011, \aap, 527, LL6   

\bibitem[Pei(1992)]{1992ApJ...395..130P} Pei, Y.~C.\ 1992, \apj, 395, 130 

\bibitem[Peterson(2001)]{2001sac..conf....3P} Peterson, B.~M.\ 2001, Advanced Lectures on the Starburst-AGN, 3 

\bibitem[Phillips et al.(2013)]{2013ApJ...779...38P} Phillips, M.~M., Simon, J.~D., Morrell, N., et al.\ 2013, \apj, 779, 38 

\bibitem[Phinney(1989)]{1989IAUS..136..543P} Phinney, E.~S.\ 1989, The Center of the Galaxy, 136, 543 

\bibitem[Poon et al.(2011)]{2011arXiv1109.0899P} Poon, H., Pun, J.~C.~S., Lam, T.~Y., Qiu, Y.~L., \& Wei, J.~Y.\ 2011, arXiv:1109.0899   

\bibitem[Poznanski et al.(2009)]{2009ApJ...694.1067P} Poznanski, D., Butler, N., Filippenko, A.~V., et al.\ 2009, \apj, 694, 1067   

\bibitem[Poznanski et al.(2011)]{2011MNRAS.415L..81P} Poznanski, D., Ganeshalingam, M., Silverman, J.~M.,   \& Filippenko, A.~V.\ 2011, \mnras, 415, L81   

\bibitem[Poznanski et al.(2012)]{2012MNRAS.426.1465P} Poznanski, D., Prochaska, J.~X., \& Bloom, J.~S.\ 2012, \mnras, 426, 1465   

\bibitem[Prieto et al.(2013)]{2013ApJ...763L..27P} Prieto, J.~L., Brimacombe, J., Drake, A.~J., \& Howerton, S.\ 2013, \apjl, 763, L27 

\bibitem[Quimby(2006)]{2006PhDT........13Q} Quimby, R.~M.\ 2006, Ph.D.~Thesis,  

\bibitem[Rees(1988)]{1988Natur.333..523R} Rees, M.~J.\ 1988, \nat, 333, 523 

\bibitem[Reilly et al.(2017)]{2017arXiv170108885R} Reilly, E., Maund, J.~R., Baade, D., et al.\ 2017, arXiv:1701.08885 

\bibitem[Richmond et al.(1994)]{1994AJ....107.1022R} Richmond, M.~W., Treffers, R.~R., Filippenko, A.~V., et al.\ 1994, \aj, 107, 1022 

\bibitem[Riess et al.(2005)]{2005ApJ...627..579R} Riess, A.~G., Li, W., Stetson, P.~B., et al.\ 2005, \apj, 627, 579   

\bibitem[Ryder et al.(2004)]{2004MNRAS.349.1093R} Ryder, S.~D., Sadler, E.~M., Subrahmanyan, R., et al.\ 2004, \mnras, 349, 1093 

\bibitem[Sana et al.(2012)]{2012Sci...337..444S} Sana, H., de Mink, S.~E., de Koter, A., et al.\ 2012, Science, 337, 444 

\bibitem[Schlafly \& Finkbeiner(2011)]{2011ApJ...737..103S} Schlafly, E.~F., \& Finkbeiner, D.~P.\ 2011, \apj, 737, 103 

\bibitem[Schlegel(1990)]{1990MNRAS.244..269S} Schlegel, E.~M.\ 1990, \mnras, 244, 269  

\bibitem[Schlegel(1996)]{1996AJ....111.1660S} Schlegel, E.~M.\ 1996, \aj, 111, 1660   

\bibitem[Schmidt et al.(1992)]{1992AJ....104.1563S} Schmidt, G.~D., Elston, R., \& Lupie, O.~L.\ 1992, \aj, 104, 1563   

\bibitem[Schmidt et al.(1992)]{1992ApJ...398L..57S} Schmidt, G.~D., Stockman, H.~S., \& Smith, P.~S.\ 1992, \apjl, 398, L57   

\bibitem[Serkowski et al.(1975)]{1975ApJ...196..261S} Serkowski, K., Mathewson, D.~S., \& Ford, V.~L.\ 1975, \apj, 196, 261   

\bibitem[Shivvers et al.(2015)]{2015ApJ...806..213S} Shivvers, I., Groh, J.~H., Mauerhan, J.~C., et al.\ 2015, \apj, 806, 213   

\bibitem[Smartt(2009)]{2009ARA&A..47...63S} Smartt, S.~J.\ 2009, \araa, 47, 63 

\bibitem[Smith(2010)]{2010MNRAS.402..145S} Smith, N.\ 2010, \mnras, 402, 145

\bibitem[Smith(2011)]{2011MNRAS.415.2020S} Smith, N.\ 2011, \mnras, 415, 2020 

\bibitem[Smith(2014)]{2014ARA&A..52..487S} Smith, N.\ 2014, \araa, 52, 487 

\bibitem[Smith \& Arnett(2014)]{2014ApJ...785...82S} Smith, N., \& Arnett, W.~D.\ 2014, \apj, 785, 82 

\bibitem[Smith, Ginsburg, \& Bally (2017)]{SmithALMAsubmitted} Smith, N., Ginsburg, A., \& Bally, J.\ 2017, submitted

\bibitem[Smith et al.(2008)]{2008ApJ...686..467S} Smith, N., Chornock, R., Li, W., et al.\ 2008, \apj, 686, 467-484 

\bibitem[Smith et al.(2010)]{2010ApJ...709..856S} Smith, N., Chornock, R., Silverman, J.~M., Filippenko, A.~V., \& Foley, R.~J.\ 2010, \apj, 709, 856   

\bibitem[Smith et al.(2008)]{2008ApJ...680..568S} Smith, N., Foley, R.~J., \& Filippenko, A.~V.\ 2008, \apj, 680, 568

\bibitem[Smith et al.(2011)]{2011MNRAS.412.1522S} Smith, N., Li, W., Filippenko, A.~V., \& Chornock, R.\ 2011, \mnras, 412, 1522   

\bibitem[Smith et al.(2015)]{2015MNRAS.449.1876S} Smith, N., Mauerhan, J.~C., Cenko, S.~B., et al.\ 2015, \mnras, 449, 1876   

\bibitem[Smith et al.(2014)]{2014MNRAS.438.1191S} Smith, N., Mauerhan, J.~C., \& Prieto, J.~L.\ 2014, \mnras, 438, 1191 

\bibitem[Smith et al.(2009)]{2009ApJ...695.1334S} Smith, N., Silverman, J.~M., Chornock, R., et al.\ 2009, \apj, 695, 1334 

\bibitem[Smith et al.(2012)]{2012AJ....143...17S} Smith, N., Silverman, J.~M., Filippenko, A.~V., et al.\ 2012, \aj, 143, 17   

\bibitem[Smith \& Tombleson(2015)]{2015MNRAS.447..598S} Smith, N., \& Tombleson, R.\ 2015, \mnras, 447, 598 

\bibitem[Smith et al.(2005)]{2005ApJ...635L..41S} Smith, N., Zhekov, S.~A., Heng, K., et al.\ 2005, \apjl, 635, L41 	

\bibitem[Stetson(1987)]{1987PASP...99..191S} Stetson, P.~B.\ 1987, \pasp, 99, 191 

\bibitem[Stritzinger et al.(2012)]{2012ApJ...756..173S} Stritzinger, M., Taddia, F., Fransson, C., et al.\ 2012, \apj, 756, 173   

\bibitem[Strubbe \& Quataert(2009)]{2009MNRAS.400.2070S} Strubbe, L.~E., \& Quataert, E.\ 2009, \mnras, 400, 2070 

\bibitem[Sugawara et al.(2015)]{2015PASJ...67..121S} Sugawara, Y., Maeda, Y., Tsuboi, Y., et al.\ 2015, \pasj, 67, 121 

\bibitem[Turatto et al.(2003)]{2003fthp.conf..200T} Turatto, M., Benetti, S., \& Cappellaro, E.\ 2003, From Twilight to Highlight: The Physics of Supernovae, 200 

\bibitem[Tuthill et al.(1999)]{1999Natur.398..487T} Tuthill, P.~G., Monnier, J.~D., \& Danchi, W.~C.\ 1999, \nat, 398, 487 

\bibitem[Ulrich et al.(1997)]{1997ARA&A..35..445U} Ulrich, M.-H., Maraschi, L., \& Urry, C.~M.\ 1997, \araa, 35, 445 

\bibitem[Valenti et al.(2016)]{2016MNRAS.459.3939V} Valenti, S., Howell, D.~A., Stritzinger, M.~D., et al.\ 2016, \mnras, 459, 3939 

\bibitem[Vinko et al.(2012)]{2012CBET.3022....1V} Vinko, J., Zheng, W., Marion, G.~H., et al.\ 2012, Central Bureau Electronic Telegrams, 3022, 1  

\bibitem[Wang et al.(2001)]{2001ApJ...550.1030W} Wang, L., Howell, D.~A., H{\"o}flich, P., \& Wheeler, J.~C.\ 2001, \apj, 550, 1030 

\bibitem[Wang et al.(2002)]{2002ApJ...579..671W} Wang, L., Wheeler, J.~C., H{\"o}flich, P., et al.\ 2002, \apj, 579, 671 

\bibitem[Wang \& Wheeler(2008)]{2008ARA&A..46..433W} Wang, L., \& Wheeler, J.~C.\ 2008, \araa, 46, 433 

\bibitem[Wardle \& Kronberg(1974)]{1974ApJ...194..249W} Wardle, J.~F.~C., \& Kronberg, P.~P.\ 1974, \apj, 194, 249 

\bibitem[Wheeler et al.(2002)]{2002ApJ...568..807W} Wheeler, J.~C., Meier, D.~L., \& Wilson, J.~R.\ 2002, \apj, 568, 807 

\bibitem[Williams et. al.(2017, in prep.)]{inprep} G. G. Williams, J. L. Hoffman, N. Smith, et al. in preparation

\bibitem[Yuan \& Akerlof(2008)]{2008ApJ...677..808Y} Yuan, F., \& Akerlof, C.~W.\ 2008, \apj, 677, 808 

\bibitem[Zhekov et al.(2014a)]{2014MNRAS.445.1663Z} Zhekov, S.~A., Tomov, T., Gawronski, M.~P., et al.\ 2014, \mnras

\bibitem[Zhekov et al.(2014b)]{2014ApJ...785....8Z} Zhekov, S.~A., Gagn{\'e}, M., \& Skinner, S.~L.\ 2014, \apj, 785, 8 
\end{thebibliography}
\end{document}